\documentclass[12pt,twoside]{article}
\usepackage[mathscr]{eucal}
\usepackage{amsmath,amsfonts,amssymb,amsthm}
\usepackage{times}
\usepackage{multibox}

\def\Q#1#2{\frac{\partial #1}{\partial #2}}

\def\QS#1#2{\frac{\partial^S #1}{\partial #2}}
\def\varQ#1#2{\frac{\delta #1}{\delta #2}}

\def\eps{\epsilon}

\def\d_V{\text{d}_V}
\def\dv{\text{d}_V}
\def\dH{\text{d}_H}

\def\d_Vphi{\text{d}_V\hspace{-0.06em}\phi}
\def\d_Vphibar{\text{d}_V\hspace{-0.06em}\bar\phi}
\def\d_Vxi{\text{d}_V\hspace{-0.06em}\xi}

\def\ndelta{\delta\hspace{-0.50em}\slash\hspace{-0.05em} }

\def\be{\begin{eqnarray}}
\def\ee{\end{eqnarray}}
\def\beann{\begin{eqnarray*}}
\def\eeann{\end{eqnarray*}}
\def\beq{\begin{equation}}
\def\eeq{\end{equation}}
\def\ba{\begin{array}}
\def\ea{\end{array}}
\def\ben{\begin{enumerate}}
\def\een{\end{enumerate}}
\def\bea{\begin{eqnarray}}
\def\eea{\end{eqnarray}}

\def\5{\bar }
\def\6{\partial }
\def\7{\hat }
\def\4{\tilde }
\advance\textheight\topskip
\renewcommand{\tilde}{\widetilde}
\renewcommand{\hat}{\widehat}
\newtheorem{prop}{Proposition}
\newtheorem{lemma}[prop]{Lemma}

\newtheorem{corollary}[prop]{Corollary}
\newtheorem{theorem}[prop]{Theorem}

\newcommand{\bref}[1]{\textbf{\ref{#1}}}

\newcommand{\dd}{\partial}
\renewcommand{\d}{\partial}

\renewcommand{\geq}{\,{\geqslant}\,}
\renewcommand{\leq}{\,{\leqslant}\,}

\newcommand{\binner}[2]{%
  {\langle}\kern-4.15pt{\langle}#1{,}\,#2{\rangle}\kern-4.15pt{\rangle}}

\newcommand{\half}{\mathchoice{%
    \ffrac{1}{2}}{\frac{1}{2}}{\frac{1}{2}}{\frac{1}{2}}}

\newcommand{\ffrac}[2]{\raisebox{.5pt}%
  {\footnotesize$\displaystyle\frac{#1}{#2}$}\kern1pt}

\newcommand{\ddl}[2]{\ffrac{\dd #1}{\dd #2}}

\newcommand{\vdl}[1]{\ffrac{{\delta}}{\delta #1}}

\newcommand{\vddl}[2]{{\ffrac{\delta #1}{\delta #2}}}

\def\cA{\mathcal{A}}

\def\cC{\mathcal{C}}

\def\cE{\mathcal{E}}
\def\cF{\mathcal{F}}

\def\cK{\mathcal{K}}
\def\cL{\mathcal{L}}

\def\cN{\mathcal{N}}

\def\cQ{\mathcal{Q}}

\def\cT{\mathcal{T}}

\def\cV{\mathcal{V}}

\numberwithin{equation}{section} \makeatletter
\@addtoreset{equation}{section}

\begin{document}

\def\mytitle{Surface charge algebra in gauge theories and
  thermodynamic integrability}

\pagestyle{myheadings} \markboth{\textsc{\small Barnich, Comp\`ere}}{%
  \textsc{\small Surface charge algebra in gauge theories:
    Integrability}} \addtolength{\headsep}{4pt}

\begin{flushright}\small
ULB-TH/06-30
\end{flushright}

\begin{centering}

  \vspace{1cm}

  \textbf{\Large{\mytitle}}

  \vspace{1cm}

  {\large Glenn Barnich$^{a}$ and Geoffrey Comp\`ere$^{b}$ }

\vspace{1cm}

\begin{minipage}{.9\textwidth}\small \it \begin{center}
   Physique Th\'eorique et Math\'ematique, Universit\'e Libre de
   Bruxelles\\ and \\ International Solvay Institutes, \\ Campus
   Plaine C.P. 231, B-1050 Bruxelles, Belgium \end{center}
\end{minipage}

\end{centering}

\vspace{.5cm}

\begin{center}
  \begin{minipage}{.9\textwidth}
    \textsc{Abstract}. Surface charges and their algebra in
    interacting Lagrangian gauge field theories are constructed out of the underlying linearized theory 
    using techniques from the variational calculus. 
    In the case of exact solutions and symmetries, the surface charges are
    interpreted as a Pfaff system. Integrability is governed by
    Frobenius' theorem and the charges associated with the derived
    symmetry algebra are shown to vanish. Surfaces charges reproduce well-known Hamiltonian and covariant phase space expressions. In the asymptotic context, we provide a generalized covariant derivation of the result that
    the representation of the asymptotic symmetry algebra through
    charges may be centrally extended. Comparision with Hamiltonian and covariant phase space methods are done in appendix.
  \end{minipage}
\end{center}

\begin{minipage}{.9\textwidth}

\end{minipage}

\vfill

\noindent
\mbox{}
\raisebox{-3\baselineskip}{%
  \parbox{\textwidth}{\mbox{}\hrulefill\\[-4pt]}}
{\scriptsize$^a$Senior Research Associate of the Fund for
  Scientific Research-FNRS (Belgium).\\ $^b$Research Fellow of the 
  Fund for Scientific Research-FNRS (Belgium). }

\thispagestyle{empty}
\newpage

\begin{small}
{\addtolength{\parskip}{-1.5pt}
 \tableofcontents}
\end{small}
\newpage

\section{Introduction}
\label{sec:introduction}

Surface charges, both in general relativity and in Yang-Mills type
gauge theories, have been extensively studied because of the central
r\^ole symmetries and conserved charges play in analyzing the
dynamics. In the case of general relativity for instance, these
charges describe total energy and angular momentum (see
e.g.~\cite{Abbott:1981ff}), while for gauge theories of Yang-Mills
type, electric, color \cite{Abbott:1982jh}, or even magnetic
charges \cite{Barnich:2007uu} may be encoded in this way.

In the Hamiltonian formulation of general relativity, these surface
charges appeared originally in the ADM approach
(see~\cite{Arnowitt:1962aa} and references therein). In this context,
the general theory of such charges and their algebra was developed in
\cite{Regge:1974zd,Henneaux:1985tv,Brown:1986ed,Brown:1986nw}.

In the Lagrangian framework, a variety of approaches have been
proposed: pseudo-tensors \cite{LL}, Komar integrals
\cite{Komar:1958wp}, Noether's method applied to linearized field
equations \cite{Katz:1996nr,Deser:2002rt,Deser:2002jk}, a quasi-local
approach \cite{Brown:1993br,Brown:2000dz} and covariant phase space
methods
\cite{Kijowski:1976ze,Crnkovic:1986ex,Ashtekar:1990gc,Lee:1990nz,Wald:1990ic,%
Wald:1993nt,Iyer:1994ys,Wald:1999wa}. Recent related work can be found for
instance in
\cite{Koga:2001vq,Anco:2001gk,Anco:2001gm,Julia:1998ys,Silva:1998ii,%
Julia:2000er,Julia:2002df,Blagojevic:2004np,Hollands:2005wt,%
Papadimitriou:2005ii}. 

Two categories of conservation laws associated with local symmetries in gauge theories can be distinguished: exact conservation laws associated with families of symmetric solutions, and asymptotic conservation laws where symmetries are defined ``close to infinity''. The underlying idea of the framework developed hereafter is that, in both cases, the surface charges and their properties are rooted in the linearized theory. 

The starting point for this work is the result \cite{Barnich:1994db,Barnich:2000zw} that exact conserved surface charges
are classified by so-called reducibility parameters, i.e. parameters of symmetries of field configurations. For general relativity for instance, this means that there is no ``universal'' non trivial surface charge because no non trivial Killing vector field exists for a generic metric. Rather, for gravity linearized around a background solution, surface charges are classified by the Killing vectors of the background \cite{Anderson:1996sc,Barnich:2004ts}.

More precisely, equivalence classes of reducibility parameters have
been shown \cite{Barnich:2000zw} to correspond to characteristic
cohomology classes in degree $n-2$, i.e., cohomology classes of the
exterior spacetime differential pulled back to the space of
solutions. If $f$ denote the reducibility parameters, the
representatives $k^{n-2}_f$ of characteristic cohomology are
constructed from the Euler-Lagrange derivatives of the Lagrangian
\cite{Barnich:2001jy}. Surface charges are then defined by 
integration of this representative over a closed $n-2$ dimensional hypersurface. 

These surface charges, which are linear in perturbations around a background field, reproduce the familiar results obtained in general relativity and Yang-Mills theories~\cite{Abbott:1981ff,Abbott:1982jh} and, in the first order formalism, reduce  identically to the Hamiltonian surface term that should be integrated and added to the constraints in order to obtain a well-defined Hamiltonian generator~\cite{Regge:1974zd,Henneaux:1985tv}.

In the full non linear theory, the surface charges of the linearized
theory can be re-interpreted as 1-forms in field space, the
appropiate mathematical framework being the variational bicomplex
associated with a set of Euler-Lagrange equations (see
e.g.~\cite{Andersonbook,Anderson1991,Olver:1993} and references
therein). These surface charge 1-forms can be used both in the exact
\cite{Wald:1993nt,Iyer:1994ys,Barnich:2003xg,Barnich:2004uw,Barnich:2005kq}
and in the asymptotic context \cite{Anderson:1996sc,Barnich:2001jy}.

In the case of exact solutions and symmetries, surface charges in the full theory are constructed by integrating the surface charge 1-forms of the linearized theory along a path in the space of symmetric configurations. The surface charge then depends only on the choice of solution, reducibility parameter, and on the homology class of the $n-2$-dimensional hypersurface. For configuration spaces of trivial topology, independence on the path holds if and only if an integrability condition is satisfied. As a new result, we prove the following theorem: under appropriate conditions, the surface charges associated with elements of the derived Lie algebra of the algebra of reducibility parameters vanish so that the surface charges represent the abelian quotient algebra of exact symmetries modulo its derived algebra. We also clarify in what sense integrability is governed by Frobenius' theorem. An important new feature of the present analysis is that we allow the reducibility parameters to vary along the field configurations. 

Since the surface charges are defined from the Euler-Lagrange equations and from an on-shell vanishing Noether current, they do not depend on total divergences that are added to the Lagrangian nor on total divergences that may be added to the Noether current. From the outset, our approach thus allows us to control the ambiguities inherent in covariant phase space methods \cite{Iyer:1994ys}.

In the asymptotic context, we elaborate on the properties of asymptotically conserved charges constructed from the linearized theory. The methods developed hereafter generalizes the analysis initiated in \cite{Barnich:2001jy} by removing the assumption of ``asymptotic linearity'' and generalizes also the proposal of \cite{Julia:2002df} in first order theories to the case of Lagrangians depending on an arbitrary number of derivatives. 

More precisely, we derive a Lagrangian version of the Hamiltonian results of \cite{Brown:1986ed,Brown:1986nw}: in the
integrable case, a suitably defined covariant Poisson bracket algebra of charges forms a centrally extended representation of the asymptotic symmetry algebra. Besides dealing with ambiguities, it also generalizes to generic gauge systems the original Lagrangian derivation of \cite{Koga:2001vq} in the context of covariant phase
space methods for diffeomorphism invariant theories.

The paper is organized as follows. We begin by recalling how central
charges appear in the context of Noether charges associated with 
global symmetries. 

Next, we fix our description of irreducible gauge theories and recall
that Noether currents associated with gauge symmetries can be chosen
to vanish on-shell. The surface charge 1-forms are defined and
their relation to what we call the invariant presymplectic $(n-1,2)$
form is established. In order to be self-contained, some results
of \cite{Barnich:2001jy} are rederived, independently of
BRST cohomological methods: surface charge 1-forms associated with
reducibility parameters are conserved, reducibility parameters form a
Lie algebra, the symmetry algebra in the present context, and,
on-shell, the covariant Poisson algebra of surface charge $1$-forms
is a representation of the symmetry algebra.

For variations that preserve the symmetries, we then show that the
charges associated with commutators of symmetries vanish.  The
integrability conditions for the surface charge 1-forms are
discussed next. In the context of the covariant phase space approach
to diffeomorphism invariant theories, they have been originally
discussed for a surface charge 1-form associated with a fixed vector
field \cite{Wald:1999wa}. Here we point out that for a given set of
gauge fields and gauge parameters, the surface charge 1-forms should
be considered as a Pfaff system and that integrability is governed by
Frobenius' theorem. This gives the whole subject a thermodynamical
flavor, which we emphasize by our notation $\ndelta \cQ_f[\dv\phi]$
for surface charge 1-forms. In the integrable case, finite charges are
defined by integrating the surface charge 1-forms along a path
starting from a fixed solution.

In the asymptotic context, we define a space of allowed fields and
gauge parameters with respect to a closed surface $S$. Asymptotic
symmetries at $S$ are defined as the quotient space of allowed gauge
parameters by ``proper'' gauge parameters associated with vanishing
charges. We prove that asymptotic symmetries form a Lie subalgebra of
the Lie algebra of all gauge parameters and show that the
representation of this algebra by a covariant Poisson bracket for the
associated conserved charges may be centrally extended. 

In Appendix \bref{sec:elements}, we give elementary definitions, fix
notations and conventions and recall the relevant formulae from the
variational bicomplex. In particular, we prove crucial properties of
the invariant presymplectic $(n-1,2)$ form associated with the
Euler-Lagrange equations of motion. Appendix \bref{appb} is devoted to
establishing the covariance of surface charge $1$-forms, while the key
result needed in order to show how integrability implies the algebra
of charges is established in Appendix~\bref{appc}. We further motivate
our Lagrangian approach in Appendix~\bref{sec:hamilt-form} by applying
it to the case of a first order Hamiltonian action and recovering
well-known results of the Hamiltonian approach. Finally, in
Appendix~\bref{app:E}, we apply our approach in the context of pure
gravity and highlight the differences with the expressions of
covariant phase space methods in the asymptotic context.

\section{Global symmetries, Noether charges and their algebra}
\label{sec:glob-symm-charge}

In a Lagrangian field theory, the dynamics is generated from the
Lagrangian $n$-form $\cL=L\,d^nx$ through the Euler-Lagrange
equations of motion
\begin{equation}
  \label{eq:34}
  \vddl{L}{\phi^i}=0. 
\end{equation}
A global symmetry $X$ is required to satisfy the condition $\delta_X
\cL=\dH k_X$. The Noether current $j_X$ is then defined through the
relation
\begin{eqnarray}
X^i\vddl{\cL}{\phi^i}=\dH j_X,\label{eq:31}
\end{eqnarray}
a particular solution of which is $j_X=k_X-I^n_X(\cL)$. The operator 
\[I^n_X(\cL)=(X^i\frac{\d L}{\d\phi^i_\mu}+\dots)(d^{n-1}x)_\mu,\] is defined
by equation \eqref{phihomotopy} for Lagrangians depending on more than
first order derivatives. Applying $\delta_{X_1}$ to the definition of
the Noether current for $X_2$ and using \eqref{eq:32} together with
the facts that ${X_1}$ is a global symmetry and that Euler Lagrange
derivatives annihilate $d_H$ exact $n$ forms, we get
\begin{eqnarray}
  \label{eq:5}
  d_H\Big(\delta_{X_1}j_{X_2}-j_{[X_1,X_2]}
-T_{X_1}[X_2,\vddl{\cL}{\phi}]\Big)=0,
\end{eqnarray}
with $T_{X_1}[X_2,\vddl{\cL}{\phi}]$ linear and homogeneous in the
Euler-Lagrange derivatives of the Lagrangian and defined in
\eqref{eq:27}. If the expression in parenthesis on the LHS of
\eqref{eq:5} is $\dH$ exact, we get 
\begin{equation}
\delta_{X_1}j_{X_2} \approx j_{[X_1,X_2]} + \dH(\cdot
),\label{alg_cur}
\end{equation}
where $\approx 0$ means equal for all solutions of the Euler-Lagrange
equations of motion. Upon integration over closed $n-1$ dimensional
surfaces, this yields the usual algebra of Noether charges when 
evaluated on solutions. 

The origin of classical central charges in the context of Noether
charges associated with global symmetries are the obstructions for the
expression in parenthesis on the LHS of \eqref{eq:5} to be $\dH$
exact, i.e., the cohomology of $\dH$ in the space of local forms of
degree $n-1$. This cohomology is isomorphic to the Rham cohomology in
degree $n-1$ of the fiber bundle of fields (local coordinates
$\phi^i$) over the base space $M$ (local coordinates $x^\mu$), see
e.g.~\cite{Andersonbook,Anderson1991}. The case of classical
Hamiltonian mechanics, $n=1$, $\cL=(p\dot q-H)dt$ is discussed for
instance in \cite{Arnoldbook}. Examples in higher dimensions can be
found in \cite{deAzcarraga:1989gm}.

\section{Gauge symmetries, surface charges and their algebra}
\label{sec:local-symm-charge}

\subsection{Gauge symmetries}
\label{sec:gauge-symmetries}

In order to describe gauge theories, one needs besides the fields
$\phi^i(x)$ gauge parameters $f^\alpha(x)$. Instead of considering the
gauge parameters as additional arbitrary functions of $x$, it is
useful to extend the jet-bundle. Because we want to consider
commutation relations involving gauge symmetries, several copies
$f^\alpha_{a(\mu)}$, $a=1,2,3\dots$, of the jet-coordinates associated
with gauge parameters are needed\footnote{Alternatively, one could
  make the coordinates $f^\alpha_{(\mu)}$ Grassmann odd, but we will
  not do so here.}. We will denote the whole set of fields as
$\Phi^\Delta_a=(\phi^i,f^\alpha_a)$ and extend the variational
bicomplex to this set. More precisely we continue to denote by $\dv$
the vertical differential that also involves the $f^\alpha_a$, while
$\dv^\phi$ denotes the part that acts on the fields $\phi^i$ and their
derivatives alone.

Let $\delta_{R_f}\phi^i=R^i_f$ be characteristics that depend linearly
and homogeneously on the new jet-coordinates $f^\alpha_{(\mu)}$,
$R^i_f=R^{i(\mu)}_\alpha f^\alpha_{(\mu)}$. We assume that these
characteristics define a generating set of gauge symmetries of
$\cL$. This means that they define symmetries and that every other
symmetry $Q_f$ that depends linearly and homogeneously on an arbitrary
gauge parameter $f$ is given by
$Q^i_f=R^{i(\mu)}_\alpha\partial_{(\mu)}
Z^\alpha_f+M^{+i}_f[\vddl{L}{\phi}]$ with
$Z^\alpha_f=Z^{\alpha(\nu)}f_{(\nu)}$ and
$M^{+i}_f[\vddl{L}{\phi}]=(-\partial)_{(\mu)}\Big(M^{[j(\nu)i(\mu)]}
\6_{(\nu)}\vddl{L}{\phi^j} f\Big)$, see
e.g.~\cite{Henneaux:1992ig,Barnich:2000zw} for more details. For
simplicity, we assume in addition that the generating set is
irreducible: if $R^{i(\mu)}_\alpha\partial_{(\mu)} Z^\alpha_f\approx
0$, then $Z^\alpha_f\approx 0$. Our results can easily be extended to
the reducible case, see e.g.~\cite{Compere:2007vx} for a recent
application. 

For all collections of local functions $Q_i$, let us define
\bea \forall
Q_i,f^\alpha:\quad R^i_fQ_i = R^{+i}_\alpha(Q_i)+\partial_\mu S^{\mu
  i}_\alpha(f^\alpha,Q_i),
\label{1.3}\\
M^{+i}[\vddl{L}{\phi}]Q_i=M^{[j(\nu)i(\mu)]}
\6_{(\nu)}\vddl{L}{\phi^j} \partial_{(\mu)}Q_i+\partial_\mu M^{\mu
  ji}(\vddl{L}{\phi^i},\vddl{L}{\phi^j}).  \eea 
If $Q_i=\vddl{L}{\phi^i}$ we get, on account of the Noether identities 
$R^{+i}_\alpha(\vddl{L}{\phi^i})=0$ and the skew-symmetry of
$M^{[j(\nu)i(\mu)]}_\alpha$, that the Noether current for a gauge
symmetry can be chosen to vanish weakly, \bea
R^i_f\vddl{\cL}{\phi^i}= \dH S_f,\label{sec1cur}\quad
M^{+i}[\vddl{L}{\phi}]\vddl{\cL}{\phi^i}= \dH M, \eea where $S_f
=S^{\mu i}_\alpha(\vddl{L}{\phi^i},f^\alpha)(d^{n-1}x)_\mu$ and
$M=M^{\mu ji}(\vddl{L}{\phi^j},\vddl{L}{\phi^i})(d^{n-1}x)_\mu$.

In the simple case where the gauge transformations depend at most on
the derivatives of the gauge parameter to first order, $R_f^i =
R_\alpha^i f^\alpha + R^{i\mu}_\alpha \d_\mu f^\alpha$, the weakly
vanishing Noether current reduces to
\begin{equation}
S_f = R^{i\mu}_\alpha f^\alpha
\vddl{L}{\phi^i}(d^{n-1}x)_\mu.\label{eq:S_first}
\end{equation}

\subsection{Surface charge 1-forms}
\label{sec:surface-charges}

Motivated by the cohomological results of \cite{Barnich:2001jy}
summarized in the introduction, we consider the $(n-2,1)$
forms~\footnote{For convenience, these forms have been defined with an
  overall minus sign as compared to the definition used
  in~\cite{Barnich:2001jy}.}
\begin{eqnarray}
k_f[\dv\phi]=I^{n-1}_{\dv \phi} S_f,
\label{def}
\end{eqnarray}
where the horizontal homotopy operator $I^{n-1}_{\dv \phi}$ is defined
in \eqref{phihomotopy}. 

For first order theories and for gauge transformations depending at
most on the first derivative of gauge parameters, the forms
$k_f[\dv\phi]$ reduce to those proposed
in~\cite{Silva:1998ii,Julia:2002df},
\begin{equation}
k_f[\dv \phi] = \half \dv \phi^i \QS{}{\phi^i_\nu}\left(
\Q{}{dx^\nu}S_f \right),
\end{equation}
with $S_f$ given in~\eqref{eq:S_first}.

The forms $k_f[\dv \phi]$ are intimately related to the invariant
presymplectic $(n-1,2)$ form $W_{{\delta\cL}/{\delta\phi}}=- \half
I^n_{\dv\phi}\big(\dv\phi^i\vddl{\cL}{\phi^i}\big)$, discussed in
details in Appendix~\ref{sec:prop-w_vddl-1}. Let $i_Q = \d_{(\mu)}Q^i
\QS{}{\dv\phi^i_{(\mu)}}$ denote contraction with 
$\delta_Q$ and
$W_{{\delta\cL}/{\delta\phi}}[\dv\phi,R_{f}]=-i_{R_f}W_{{\delta\cL}/{\delta\phi}}$.

\begin{lemma}\label{lemma_ch} The forms $k_f[\dv\phi]$
  satisfy
\begin{eqnarray}
\dH k_f[\dv\phi]=W_{{\delta\cL}/{\delta\phi}}[\dv\phi,R_{f}] -\dv^\phi
S_f +T_{R_f}[\dv\phi,\vddl{\cL}{\phi}] ,\label{eq:29a}
\end{eqnarray}
where $T_{R_f}[\dv\phi,\vddl{\cL}{\phi}]$, defined explicitly in
\eqref{eq:27}, vanishes on-shell.
\end{lemma}

\vspace*{.25cm}

\begin{minipage}{.90\textwidth}\footnotesize

\noindent Indeed, it follows from \eqref{sec1cur} and \eqref{eq:36} that
\begin{eqnarray}
I^n_{\dv\phi} (\dH
S_{f})=W_{{\delta\cL}/{\delta\phi}}[\dv\phi,R_{f}]+
T_{R_{f}}[\dv\phi,\vddl{\cL}{\phi}].\label{eq:40}
\end{eqnarray}
The result follows by combining with equation \eqref{cc1a}. \qed 

\end{minipage}

\vspace*{.25cm}

\noindent We will consider 1-forms $\dv^s\phi$ that are tangent to the
space of solutions. These 1-forms are to be contracted with
characteristics $Q_s$ such that $\delta_{Q_s}\vddl{L}{\phi^i}\approx
0$.  In particular, they can be contracted with characteristics $Q_s$
that define symmetries, gauge or global, since
$\delta_{Q_s}\cL=\dH(\cdot)$ implies
$\delta_{Q_s}\vddl{L}{\phi^i}\approx 0$ on account of \eqref{eA7} and
\eqref{eq:3}.  For such 1-forms, 
\begin{eqnarray}
\dH k_f[\dv^s\phi] \approx
W_{{\delta\cL}/{\delta\phi}}[\dv^s\phi,R_{f}].\label{eq:29abis}
\end{eqnarray}
Applying the homotopy operators $I_f^{n-1}$ defined in~\eqref{eq:45} to
\eqref{eq:29abis}, one gets
\begin{eqnarray}
k_f[\dv^s \phi]&\approx& I_f^{n-1}
W_{{\delta\cL}/{\delta\phi}}[\dv^s\phi,R_{f}]
+\dH(\cdot).
\label{4.13}
\end{eqnarray}
Note that this relation holds off-shell,
 \begin{eqnarray}
k_f[\dv\phi]&=& I_f^{n-1}
W_{{\delta\cL}/{\delta\phi}}[\dv\phi,R_{f}] +\dH(\cdot),
\label{4.13bis}
\end{eqnarray}
if 
\begin{equation}\label{simplif}
I^{n-1}_f  S_f =0,\qquad I^{n-1}_f
T_{R_f}[\dv\phi,\vddl{\cL}{\phi}]) = 0.
\end{equation}
As we will see below, the relevant components of this condition hold
for instance for Einstein gravity and in the Hamiltonian framework.

For a given closed $n-2$ dimensional surface $S$, which we typically
take to be a sphere inside a hyperplane, the surface charge $1$-forms
are defined by
\begin{eqnarray}
\ndelta \cQ_f[\dv\phi]=\oint_{S} k_f[\dv\phi].
  \label{eq:17}
\end{eqnarray}

\begin{lemma}
The surface charge 1-forms contracted with gauge transformations are
on-shell skew-symmetric in the sense that  
\begin{eqnarray}
  \label{eq:48}
  \ndelta \cQ_{f_2}[R_{f_1}]\approx -\ndelta \cQ_{f_1}[R_{f_2}].
\end{eqnarray}
\end{lemma}
\vspace*{.25cm}

\begin{minipage}{.90\textwidth}\footnotesize

Applying $i_{R_{f_1}}$ to \eqref{eq:29a} in terms of $f_2$, and using
$I^{n-1}_{f_1}$, we get
\[k_{f_2}[R_{f_1}] 
\approx - I^{n-1}_{f_1}
W_{{\delta\cL}/{\delta\phi}}[R_{f_1},R_{f_2}]+\dH(\cdot).\] 
Comparing
with $i_{R_{f_1}}$ applied to \eqref{4.13} in terms of $f_2$, this
implies 
\begin{eqnarray}
k_{f_2}[R_{f_1}]\approx -k_{f_1}[R_{f_2}]+\dH(\cdot),\label{eq:47}
\end{eqnarray}
from which the lemma follows by integration. \qed

\end{minipage}

\vspace*{.25cm}

\noindent At a fixed solution $\phi_s$ to the Euler-Lagrange equations
of motion, we consider the space $\mathfrak e_{\phi_s}$ of gauge
parameters $f^s$ that satisfy $R^i_{f^s}|_{\phi_s}=0$.  We call such
gauge parameters exact reducibility parameters at $\phi_s$.

The surface charge $1$-forms associated with reducibility parameters
are $\dH$-closed on-shell. More precisely, equation (\ref{eq:29a})
implies that $d_H k_{f^s}[\dv^s\phi]|_{\phi_s}=0$ for 1-forms $\dv^s
\phi$ tangent to the space of solutions. As a consequence of Lemma
\bref{lemma_ch}, we then have 
\begin{corollary}\label{prop1}
  The surface charge 1-forms~$\ndelta \cQ_{f^s}[\dv^s
\phi]|_{\phi_s}$ associated with reducibility parameters
  only depend on the homology class of $S$.
\end{corollary}
In particular, if $S$ is the sphere ($t,r$ constant) in spherical
coordinates for instance, $\ndelta \cQ_{f^s}[\dv^s\phi]|_{\phi_s}$ is
$r$ and $t$ independent, and thus does not depend on any of the
coordinates, but only on the solution and the tangent vector in the
space of solutions.

{\bf Remarks:}

\begin{enumerate}

\item Trivial gauge transformations $\delta\phi^i =
M^{+i}_f[\vddl{L}{\phi}]$ can be associated with a $(n-2,1)$ form
$k_f = I_{\dv\phi}^{n-1}M_f$ in the same way as~\eqref{def} with
$M_f$ defined in~\eqref{sec1cur}. Now, $k_f \approx 0$ since
the homotopy operator~\eqref{phihomotopy} can only ``destroy'' one
of the two equations of motion contained in $M_f$. Therefore,
trivial gauge transformations are associated with on-shell vanishing 
surface charge 1-forms.

\item As briefly recalled in the introduction, one can in fact show
under suitable assumptions
\cite{Barnich:1994db,Anderson:1996sc,Barnich:2000zw,%
  Barnich:2001jy,Barnich:2004ts} that any other $(n-2,1)$ form that is
closed at a given solution $\phi_s$ when contracted with
characteristics tangent to the space of solutions differs from a form
$k_{f^s} [\dv\phi]$ associated with some reducibility $f^s$ parameters
at most by terms that are $\dH$-exact or vanish when contracted with
characteristics tangent to the space of solutions. 

\item We will show in Appendix~\ref{sec:prop-w_vddl-1} that
\begin{eqnarray}
  \label{eq:79a}
 - W_{{\delta\cL}/{\delta\phi}}=\Omega_{\cL}+\dH
E_{\cL},\qquad\dv \Omega_{\cL}=0,
\end{eqnarray}
where $\Omega_{\cL}$ is the standard presymplectic $(n-1,2)$-form used
in covariant phase space methods, and $E_{\cL}$ is a suitably defined
$(n-2,2)$ form. Contracting~\eqref{eq:79a} with a gauge transformation
$R_f$ it follows from \eqref{4.13} that, apart from on-shell and
$\dH$-exact terms, $k_f[\dv^s\phi]$ differs by the additional term
$E_{\cL}[\dv\phi,R_f]$ from similar $(n-2,1)$-forms derived in the
context of covariant phase space methods in
\cite{Lee:1990nz,Wald:1993nt,Iyer:1994ys}.
\end{enumerate}

\subsection{Algebra}
\label{sec:central-extensions}
~
Because we have assumed that $\delta_{R_f} \phi^i=R^i_f$ provide a
generating set of non trivial gauge symmetries, the commutator algebra
of the non trivial gauge symmetries closes on-shell in the sense that
\bea \delta_{R_{f_1}}R^i_{f_2}-(1\longleftrightarrow 2)
=-R^i_{[f_1,f_2]}+M^{+i}_{f_1,f_2}[\vddl{L}{\phi}] ,\label{1.18} \eea
with
$[f_1,f_2]^\gamma=C^{\gamma(\mu)(\nu)}_{\alpha\beta}f^\alpha_{1(\mu)}f_{2(\nu)}^\beta$
for some skew-symmetric functions $C^{\gamma(\mu)(\nu)}_{\alpha\beta}$
and for some characteristic $M^{+i}_{f_1,f_2}[\vddl{L}{\phi}]$. 

At any solution $\phi_s(x)$ to the Euler-Lagrange equations
of motion, the space of all gauge parameters equipped with the
bracket $[\cdot,\cdot]$ is a Lie algebra.

\vspace*{.25cm}

\begin{minipage}{.90\textwidth}\footnotesize

Indeed, by applying $\delta_{R_{f_3}}$ to \eqref{1.18} and taking
cyclic permutations, one gets $R_{[[f_1,f_2],f_3]}+{\rm cyclic}\
(1,2,3)\approx 0$ on account of $\delta_{R_f}\vddl{L}{\phi^i}\approx
0$. Irreducibility then implies the Jacobi identity
\begin{eqnarray}
  \label{eq:46}
  [[f_1,f_2],f_3]^\gamma+{\rm cyclic}\
(1,2,3)\approx 0.
\end{eqnarray}

\end{minipage}

\vspace*{.25cm}

\noindent It then also follows from \eqref{1.18} and \eqref{eq:46} that
$\mathfrak e_{\phi_s}$ is a Lie algebra, the Lie algebra of exact
reducibility parameters at the particular solution $\phi_s$.
\begin{prop}\label{lem11}
When evaluated at a solution $\phi_s$, for $1$-forms tangent to the space of
solutions and for a reducibility parameter $f^s$ at $\phi_s$, the
$(n-2,1)$-forms $k_{f}[\dv^s \phi]$ are covariant up to $\dH$ exact
terms, 
\begin{eqnarray}
  \label{eq:35}
  \delta_{R_{f_1}} k_{f^s_2}[\dv^s \phi] \approx -k_{[f_1,f^s_2]}[\dv^s
  \phi]+d_H(\cdot).
\end{eqnarray}
\end{prop}
This proposition is proved in Appendix \bref{appb}.  If the Lie
bracket of surface charge 1-forms is defined by
\begin{eqnarray}
  \label{eq:60}
  [\ndelta \cQ_{f_1},\ndelta \cQ_{f_2}]=-\delta_{R_{f_1}} \ndelta \cQ_{f_2},
\end{eqnarray}
we thus have shown:
\begin{corollary}\label{cor3}
At a given solution $\phi_s$ and for 1-forms $\dv^s\phi$ tangent to
the space of solutions, the Lie algebra of surface charge
1-forms represents the Lie algebra of exact reducibility parameters
$\mathfrak e_{\phi_s}$,
\begin{eqnarray}
  \label{eq:47b}
  [\ndelta \cQ_{f^s_1} ,\ndelta \cQ_{f^s_2}][\dv^s\phi]|_{\phi_s} =
\ndelta \cQ_{[f^s_1,f^s_2]}[\dv^s\phi]|_{\phi_s}.
\end{eqnarray}
\end{corollary}

\subsection{Exact solutions and symmetries}
\label{sec:exact-solut-symm}

Suppose one is given a family of exact solutions $\phi_s\in\cE$ 
admitting ($\phi_s$-dependent) reducibility parameters $f^s \in
\mathfrak e_{\phi_s}$, which
contains a background solution $\bar \phi$. Elements of the Lie algebra
of exact reducibility parameters $\mathfrak e_{\phi_s}$ at $\phi_s$
are denoted by $f^s$. Let us denote by $\bar \phi$ an element of this
family that we single out as the reference solution and let $\bar f
\in \mathfrak e_{\bar \phi}$ be the associated reducibility
parameters. We consider $1$-forms $\dv^sf$ that are tangent to the
space of reducibility parameters. They are to be contracted with gauge
parameters $q^s$ such that
\begin{eqnarray}
0=(\dv^{s} R_f)|_{\phi_s,f^s,Q_s,q_s}=
\delta_{Q_s}R_{f^s}|_{\phi_s}+R_{q^s}|_{\phi_s}.\label{eq:90}
\end{eqnarray}
Definition \eqref{eq:60} and Corollary \bref{cor3} imply
\begin{corollary}\label{cor4}
  For variations preserving the reducibility identities as in
  \eqref{eq:90}, the surface charge 1-forms vanish for
  elements of the derived Lie algebra ${\mathfrak e}^\prime_{\phi_s}$
  of exact reducibility parameters at $\phi_s$,
\begin{eqnarray}
  \label{eq:92}
\ndelta  \cQ_{[f^s_1,f^s_2]}[\dv^s\phi]|_{\phi_s}=0. 
\end{eqnarray}
In this case, the Lie algebra of surface charge 1-forms represents
non trivially only the abelian Lie algebra ${\mathfrak
  e}_{\phi_s}/{\mathfrak e}^\prime_{\phi_s}$.
\end{corollary}

The surface charge $Q_{\gamma_s,f}$ of
$\Phi_s=(\phi_s,f^s)$ with respect to the fixed background
$\bar\Phi=(\bar\phi,\bar f)$ is defined as
\begin{eqnarray}
  \label{eq:82}
  \cQ_{\gamma_s,f} [\Phi,\bar \Phi] = \int_{\gamma_s} 
\ndelta \cQ_{f^{s\prime}}[\dv^s\phi^\prime]
+ N_{\bar f}[\bar \phi],
\end{eqnarray}
where integration is done along a path $\gamma_s$ in the space of
exact solutions $\cE$ that joins $\bar \phi$ to $\phi_s$ for some
reducibility parameters that vary continuously along the path from
$\bar f$ to $f^s$. Note that these charges depend only on the homology
class of $S$ because equation (\ref{eq:29a}) implies that $d_H
k_{f^s}[\dv^s\phi]|_{\phi_s}=0$. Because \eqref{eq:92} holds in this
case, we have:
\begin{corollary}
  If the normalization of the background is chosen to vanish, the
  surface charges associated with elements of the derived Lie algebra
  $\mathfrak e^\prime_{\phi^s}$ of reducibility parameters vanish at
  any $\phi_s\in \cE$,
\begin{eqnarray}
  \label{eq:97}
  \cQ_{\gamma_s,[f_1,f_2]}=0.
\end{eqnarray}
\end{corollary}

The integrability conditions for the surface charges involve the
$2$-forms $\oint_S \dv^{s} k_{f^s}[\dv^s\phi]|_{\phi_s}$. Assumption
\eqref{eq:90} together with \eqref{eq:29a}, \eqref{eq:79a} imply that
$\dH \dv^{s} k_{f^s}[\dv^s\phi]|_{\phi_s}=0$, so that the
integrability conditions also only depend on the homology class of
$S$.

We now assume that solutions to the system $\vddl{L}{\phi}=0,
R^i_f=0$ are described by fields $\phi_s(x;a)$ depending smoothly on
$p$ parameters, $a^A$, $A=1,\dots p$ and reducibility parameters
$f^s(x;a,b)$ depending linearily on some additional ones $b^i$,
$i=1,\dots q$. It follows that $e_{i}(x;a)=\frac{\d}{\d b^i}
f^s(x;a,b)$ is a basis of the Lie algebra $\mathfrak e_{\phi_s}$,
$i=1,\dots r$. For each basis element $e_{ i}(x;a)$, we consider the
1-forms in parameters space $\theta_i(a,da)=\oint_S k_{e_{
    i}}[d^a\phi_s(x;a)]$, where $d^a$ is the exterior derivative in
parameter space. We thus have a Pfaff system in parameter space and
the question of integrability can be addressed using Frobenius'
theorem.

In the completely integrable case for instance, there exists an
invertible matrix $S^i_j(a)$ such that
\begin{eqnarray}
d^a f_j(a)= S^i_j(a)\theta_i(a,da)=\oint_S
k_{S^i_j(a)e_{ i}}[d^a\phi_s(x;a)]\label{eq:69}.
\end{eqnarray}
In other words, if $g_{j}(x;a)=S^i_j(a)e_{i}(x;a)$, the surfaces
charges
\begin{eqnarray}
  \label{eq:84a}
  \cQ_{j}[\Phi,\bar \Phi] = \int_{\gamma_s} 
  \ndelta\cQ_{g^\prime_{ j}}[\dv \phi^\prime] + N_{\bar g_j}[\bar \phi],
\end{eqnarray}
do not depend on the path $\gamma_s\in\cE$ but only on the final point
$(\phi_s,g_{j})$ and the initial point $(\bar\phi,\bar g_j)$.

{\bf Remarks:}

\begin{enumerate}
\item Because $d_H k_{f^s}[\dv^s\phi]|_{\phi_s}=0$ for solutions of the
source-free equations of motion, one gets the following generalization
of Gauss's law for electromagnetism in the case where the surface $S$
surrounds several sources $i$ that can be enclosed by surfaces
$S^i$:
\begin{equation}
\ndelta \cQ_{f^s}[\dv^s
\phi]|_{\phi_s} = \sum_{i}
\oint_{S^i} k_{f^s}[\dv^s\phi]|_{\phi_s},\label{eq:propG}
\end{equation}
with similar decompositions holding for the charges $\cQ_{\gamma_s,f}$ and 
$\cQ_{j}[\Phi,\bar \Phi]$. 

\item In the case of exact solutions and exact symmetries thereof, the
  theory of charges developed above does not depend on asymptotic
  properties of the fields near some boundary.

\item In the case where the surface charge is evaluated at infinity for
instance, a simplification occurs when $\Phi$ approaches $\bar \Phi$
sufficiently fast at infinity in the sense that the $(n-2,1)$-form can
be reduced to
\begin{eqnarray}
k_{f}[\dv \phi;\phi]|_{S^\infty}= k_{\bar f}[\dv \phi;\bar
\phi]|_{S^\infty}.\label{AL}
\end{eqnarray}
We call this the \emph{asymptotically linear case}. It was treated in
details in \cite{Barnich:2001jy}. In this case, the
charge~\eqref{eq:82} is manifestly path-independent and reduces to
\begin{eqnarray}
  \label{eq:82bis}
  \cQ_{f} [\Phi,\bar \Phi] =
\oint_{S^\infty} k_{\bar f}[\phi-\bar \phi;\bar \phi] + N_{\bar f
}[\bar \phi].
\end{eqnarray}
\end{enumerate}

\section{Asymptotic analysis}
\label{sec:asymptotic-analysis-1}

\subsection{Space of admissible fields and gauge parameters}
\label{sec:phasespace}

Consider for definiteness the closed surface $S^{\infty,t}$, which is
obtained as the limit when $r$ goes to infinity of the surface
$S^{r,t}$, the intersection of the space-like hyperplane $\Sigma_t$
defined by constant $t$ and the time-like or null hyperplane $\cT_r$
defined by constant $r$. The remaining angular coordinates are denoted
by $\theta^A$ and $y^a=(t,\theta^A)$. Most considerations below only
concern the region of space-time close to $S^{\infty,t}$.

We now define the space of allowed (asymptotic) solutions $\cF^s$ and
for each $\phi_s \in \cF^s$, the space of allowed gauge parameters $g
\in \cA_{\phi_s}$. They are restricted by the following requirements:

\begin{itemize}

\item The allowed gauge parameters should be such that the associated
  gauge transformations leave the allowed field configurations
  invariant, 
\begin{eqnarray}\label{eq:condRf}
  \delta_{R_g} \phi^i = R_g^i\, \text{ should be tangent to $\cF^s$}.
\end{eqnarray}
It implies that all the relations below should be valid for
$\dv^s\phi^i$ replaced by $R^i_g$.

\item Integrability of the surface charges,
\begin{eqnarray}
  \oint_{S^{r,t}} \dv^s  k_{g}[\dv^s \phi]\approx o(r^0)\,.\label{eq:92a}
\end{eqnarray}
This condition guarantees that the surface charges~\eqref{eq:82}
are independent on the path $\gamma \in \cF^s$ provided that no global
obstructions occurs in $\cF^s$. 

\item Additional conditions on $\dv E_\cL$,
\begin{eqnarray}
  \label{eq:109}
   \oint_{S^{r,t}}i_{R_g} \dv^s  E_{\cL}[\dv^s\phi,\dv^s\phi]\approx o(r^0),\; 
\oint_{S^{r,t}}\delta_{R_g} \dv^s  E_{\cL}[\dv^s\phi,\dv^s\phi]\approx o(r^0).
 \end{eqnarray}
 These assumptions are needed below in order to prove that asymptotic
 symmetries form an algebra. As we will see below, they are
 automatically fulfilled in the Hamiltonian formalism in Darboux
 coordinates.

\item Asymptotic $r$-independence of the charges,
\begin{eqnarray}
  \label{eq:105}
  \oint_{S^{r,t}} \cL_{\d_r}   k_{g}[\dv^s\phi]\approx o(r^{-1})\,,
\end{eqnarray}
where $\cL_{\d_r}$ is defined by \eqref{app:Liec}. This condition
expresses that the surface charge 1-forms \eqref{eq:17} for $S =
S^{r,t}$ are $r$-independent when $r \rightarrow \infty$. It implies
in particular finiteness of the charges. 

\item Conservation in time of the surface charges for solutions
  $\phi_s \in \cF^s$ and tangent 1-forms $\dv^s\phi$ to $\cF^s$,
\begin{eqnarray}\label{conserveda}
  \oint_{S^{r,t}} \cL_{\d_t} k_{g}[\dv^s\phi]\approx o(r^0)\,.
\end{eqnarray}

\end{itemize}

\subsection{Asymptotic symmetry algebra}
\label{sec:asympt-symm-algebra}

As a consequence of the requirements \eqref{eq:condRf}, \eqref{eq:92a}
and \eqref{eq:109}, we prove in Appendix~\bref{appc}:
\begin{prop}\label{lem4}
  For any field $\phi_s \in \cF^s$, 1-form $\dv^s\phi$ tangent to
  $\cF^s$ at $\phi_s$ and for allowed gauge parameters $g_1,g_2 \in
  \cA_{\phi_s}$, the identity
\begin{eqnarray}
  \label{eq:87}
   \oint_{S^{r,t}}k_{[g_1,g_2]}[\dv^s\phi]
 \approx \oint_{S^{r,t}}\dv^s k_{g_1}[R_{g_2}]+o(r^0)
\end{eqnarray}
holds.
\end{prop}
This allows us to show:
\begin{corollary}\label{cor:14}
  The space of allowed gauge parameters $\cA_{\phi_s}$ at $\phi_s
  \in \cF^s$ forms a Lie algebra.
\end{corollary}

\vspace*{.25cm}

\begin{minipage}{.90\textwidth}\footnotesize

  Indeed, owing to \eqref{1.18}, if $R_{g_1}, R_{g_2}$ are tangent to
  $\cF^s$ then so is $R_{[g_1,g_2]}$ and furthermore, conditions
  \eqref{eq:109} hold for $[g_1,g_2]$ if they hold for $g_1$, $g_2$
  because of relation \eqref{eq:7} and the last of \eqref{eq:comm}.
  Applying $\dv^s$ to \eqref{eq:87} implies that the charges
  associated with the parameters $[g_1,g_2]$ are integrable. Finally,
  applying $\cL_{\d_\mu}$ with $\mu = t,\, r$ to \eqref{eq:87} and
  using~\eqref{eq:condRf} shows that conditions \eqref{eq:105} and
  \eqref{conserveda} hold for $[g_1,g_2]$ if they hold for $g_1,g_2$.
  \qed

\end{minipage}

\vspace*{.25cm}

\noindent The subspace of allowed gauge parameters, $g_P \in
\cA_{\phi_s}$, satisfying
\begin{eqnarray}
  \label{eq:107}
  \oint_{S^{r,t}} k_{g_P}[\dv^s\phi] \approx o(r^0),
\end{eqnarray}
for all $\dv^s\phi$ tangent to $\cF^s$ will be called proper gauge
parameters at $\phi_s$. The associated transformations
$\delta\phi^i=R^i_{g_P}$ will be called proper gauge
transformations. On the contrary, gauge parameters (resp.
transformations) related to non vanishing surface charge 1-forms will
be called improper gauge parameters (resp.  transformations). Improper
gauge transformations send field configurations to inequivalent field
configurationsin the sense that they change the conserved charges.

Proposition~\bref{lem4} also directly implies:
\begin{corollary}\label{cor17}
Proper gauge transformations at $\phi_s \in \cF^s$ form an
ideal $\cN_{\phi_s}$ of $\cA_{\phi_s}$.
\end{corollary}

The quotient space $\cA_{\phi_s}/\cN_{\phi_s}$ is therefore a Lie
algebra which we call the asymptotic symmetry algebra $\mathfrak
e^{as}_{\phi_s}$ at $\phi_s \in \cF^s$.

{\bf Remarks:} 
\begin{enumerate}

\item Exact reducibility parameters $f^s \in \mathfrak e_{\phi_s}$
  belonging to $\cA_{\phi_s}$ survive in the asymptotic symmetry
  algebra if there exists at least one solution $\phi_s\in\cF^s$ and a
  tangent 1-form $\dv^s\phi$ such that $\ndelta \cQ_{f^s}[\dv
  \phi]|_{\phi_s} \neq 0$.

\item If the relevant components of condition~\eqref{simplif} hold and
  if the bracket of gauge transformations closes off-shell, i.e.,
  if~\eqref{1.18} hold with $M^{+i}_{f_1,f_2}[\varQ{L}{\phi}]=0$, the
  whole discussion can be done off-shell. In other words, one can
  define the space of allowed field configurations $\phi\in \cF$ and
  of allowed gauge parameters $g\in \cA_\phi$ by imposing the
  requirements of subsection~\bref{sec:phasespace} with strong instead
  of weak equalities with all results of the present subsection
  holding true for $\cF$ and $\cA_\phi$ instead of $\cF^s$ and
  $\cA_{\phi_s}$. These conditions hold for instance in the case of
  the Hamiltonian formalism discussed in
  Appendix~\bref{sec:hamilt-form} and also for Einstein
  gravity discussed in Appendix~\bref{app:E}. 

\item A way to avoid assumptions \eqref{eq:109} is to consider
instead of the $(n-2,1)$-forms $k_f[\dv\phi]$ the forms
\begin{eqnarray}
  \label{eq:23}
  k^\prime_f[\dv\phi]=k_f[\dv\phi]-E_{\cL}[\dv\phi,R_f].\label{kt}
\end{eqnarray}
Using \eqref{eq:79a}, we now have instead of
\eqref{eq:29a} and \eqref{4.13}
\begin{eqnarray}
\dH k^\prime_f[\dv\phi]= \Omega_{\cL}[R_f,\dv\phi] -\dv^\phi S_f
+T_{R_f}[\dv\phi,\vddl{\cL}{\phi}],\label{2nd}\\
k^\prime_f[\dv^s \phi]\approx I_f^{n-1}\Omega_{\cL}[R_f,\dv^s\phi]
+\dH(\cdot).
\end{eqnarray}
In the proof of Proposition \bref{lem4} in Appendx \bref{appc}, this amounts
to replacing $W_{\delta\cL/\delta\phi}$ by $-\Omega_{\cL}$ and the
additional conditions \eqref{eq:109} are not needed on account of
$\dv\Omega_{\cL}=0$.

Contrary to $k_f[\dv\phi]$ however, the forms $k^\prime_f[\dv\phi]$ depend on
the explicit choice of boundary terms in the Lagrangian. Indeed, if
$\cL \rightarrow \cL + \dH \mu$, one has $E_\cL \rightarrow E_\cL +
\dv I_{\dv \phi}\mu + \half \dH I^{n-2}_{\dv\phi}I^{n-1}_{\dv\phi}\mu$
and the resulting change in the $n-2$ form is given by
\begin{equation}
k^\prime_f[\dv\phi] \rightarrow  k^\prime_f[\dv\phi]
+\delta_{R_f}I_{\dv \phi }\mu - \dv I_{R_f}\mu - \dH(\half
i_{R_f}I^{n-2}_{\dv\phi}I^{n-1}_{\dv\phi}\mu). 
\end{equation}

\end{enumerate}

\subsection{Poisson bracket representation}
\label{sec:poiss-brack-repr}

Applying consecutively $i_{R_{g_2}}$ and $i_{R_{g_3}}$
to the integrability conditions~\eqref{eq:92a} gives
\begin{eqnarray}
\oint_{S^{\infty,t}} k_{g_1}[R_{[g_2,g_3]}] &=&
\oint_{S^{\infty,t}} \big( \delta_{R_{g_3}}k_{g_1}[R_{g_2}]- (2
\leftrightarrow 3)\big).
\end{eqnarray}
Using Proposition~\bref{lem4} on the two terms on the RHS and the
antisymmetry~\eqref{eq:47}, we get
\begin{eqnarray}
  \label{eq:89}
  \oint_{S^{\infty,t}} k_{[g_1,g_2]}[R_{g_3}]|_{\phi^s}+{\rm cyclic}\ (1,2,3)=0.
\end{eqnarray}
If we define
\begin{eqnarray}
  \label{eq:16}
  \cK_{f_1,f_2}=-\oint_S k_{f_2}[R_{f_1}]|_{\phi_s}=
\oint_{S^{\infty,t}} I^{n-1}_{f_2}
W_{{\delta\cL}/{\delta\phi}}[R_{f_1},R_{f_2}]|_{\phi_s},
\end{eqnarray}
we have shown:
\begin{corollary}\label{lem2}
$\cK_{g_1,g_2}$ defines a Chevalley-Eilenberg 2-cocycle on the
Lie algebra  $\mathfrak e^{as}_{\phi_s}$,
\begin{eqnarray}
  \label{eq:95}
  \cK_{g_1,g_2}+\cK_{g_2,g_1}&=&0,\\
 \cK_{[g_1,g_2],g_3}+ {\rm cyclic}\ (1,2,3)&=&0.
\end{eqnarray}
\end{corollary}

The surface charge $\cQ_{\gamma} [\Phi,\bar \Phi]$ of $\Phi=(\phi,f)$ with
respect to the fixed background $\bar \Phi=(\bar\phi,\bar f)$ is defined as
\begin{equation}
  \cQ_{\gamma} [\Phi,\bar \Phi] := \int_{\gamma_s} \oint_{S^{\infty,t}}
  k^\prime_{f^\prime}[\dv \phi^\prime] + N_{\bar f}[\bar \phi],
\label{def_Q}
\end{equation}
where the integration is done along a path $\gamma_s$ joining $\bar
\Phi$ to $\Phi$ in $\cF^s$. We assume here that there are no global
obstructions in $\cF^s$ for \eqref{eq:92a} to guarantee that the
surface charges
\begin{eqnarray}
  \label{eq:85}
  \cQ[\Phi,\bar \Phi]=\int_{\gamma_s}\oint_{S^{\infty,t}}
  k_{g^\prime}[\dv\phi^\prime]
+N_{\bar g}[\bar\phi],
\end{eqnarray}
do not depend on the path $\gamma\in\cF^s$. If we denote $\cQ_i \equiv
\cQ[\Phi_i,\bar \Phi_i]$ the charge related to $\Phi_i =(\phi,g_i)$,
the covariant Poisson bracket of these surface charges is defined
by
\begin{eqnarray}
  \label{eq:18}
  \{\cQ_{1} ,\cQ_{2} \}_c := -\delta_{R_{g_1}}
  \cQ_{2}=-\oint_{S^{\infty,t}}k_{g_2}[R_{g_1}].
\end{eqnarray}

For an arbitray path $\gamma_s\in\cF^s$, definitions~(\ref{eq:18}) 
and (\ref{eq:16}) lead to
\begin{eqnarray}
  \{ \cQ_{1} , \cQ_{2} \}_c- \cK_{\bar g_1,\bar g_2}[\bar \phi] &=&
  -\int_{\gamma_s}\oint_{S^{\infty,t}}  \dv^{\prime,s}
  \big(k_{g^\prime_2}[R^\prime_{g^\prime_1}])\\ &=&
\int_{\gamma_s}\oint_{S^{\infty,t}}
   k_{[g_1,g_2]}[\dv^s\phi]|_{\phi_s},
\label{equal1}
\end{eqnarray}
where proposition \bref{lem4} has been used in the last line.
Defining $\cQ_{[1,2]}$ as associated with $[g_1,g_2]$, we thus get
\begin{theorem}\label{lem5}
In $\cF^s$, the charge algebra between a fixed background solution
$\bar \phi$ and a final solution $\phi_s$ is determined by
\begin{eqnarray}
  \label{eq:71}
  \{ \cQ_{1} , \cQ_{2} \}_c=\cQ_{[1,2]}+\cK_{\bar g_1,\bar
  g_2}[\bar \phi]-N_{[\bar g_1,\bar g_2]}[\bar \phi],
\end{eqnarray}
where the central charge $\cK_{\bar g_1,\bar g_2}[\bar \phi]$ is a
2-cocycle on the Lie algebra of asymptotic symmetries $\mathfrak
e^{as}_{\bar \phi_s}$.
\end{theorem}

{\bf Remarks:}
\begin{enumerate}

\item The central charge is non-trivial if one cannot find a
      normalization $N_{\bar g}[\bar \phi]$ of the background such
      that $\cK_{\bar g_1,\bar g_2}[\bar \phi]=N_{[\bar g_1,\bar
      g_2]}[\bar \phi]$.

\item The central charge involving an exact reducibility parameter
      of the background vanishes.

\item For a semi-simple algebra $\mathfrak e^{as}_{\bar \phi_s}$,
       the property $H^2(\mathfrak e^{as}_{\bar \phi_s})=0$ guarantees
       that the central charge can be absorbed by a suitable
       normalization of the background, while the property
       $H^1(\mathfrak e^{as}_{\bar \phi_s})=0$ implies that this fixes
       the normalization completely.

\item As a consequence of Theorem~\bref{lem5} together with
       Corollary~\bref{cor17}, proper gauge transformations 
       act trivially on the charges
\begin{equation}
\delta_{R_{g_P}}\cQ_i = 0,
\end{equation}
once we assume that the normalizations associated with proper gauge
parameters all vanish.
\end{enumerate}

\section*{Acknowledgements}
\label{sec:acknowledgements}

The authors want to thank Dr.~Jun-ichirou Koga for useful
correspondence. This work is supported in part by a ``P{\^o}le
d'Attraction Interuniversitaire'' (Belgium), by IISN-Belgium,
convention 4.4505.86, by the Fund for Scientific Research-FNRS
(Belgium), by Proyectos FONDECYT 1051084, 7070183 and 1051064 (Chile)
and by the European Commission programme MRTN-CT-2004-005104, in which
the authors are associated with V.U.~Brussels.

\appendix

\section{Elements from the variational bicomplex}
\label{sec:elements}

\subsection{Elementary definitions and conventions}
\label{sec:elem-defin-conv}

We assume for notational simplicity that all fields $\phi^i$ are
(Grassmann) even.

Consider $k$-th order derivatives $\frac{\6^k\phi^i(x)}{\6
  x^{\mu_1}\dots\6 x^{\mu_k}}$ of a field $\phi^i(x)$. The
corresponding jet-coordinate is denoted by $\phi^i_{\mu_1\dots\mu_k}$.
Because the derivatives are symmetric under permutations of the
derivative indices $\mu_1,\dots,\mu_k$, these jet-coordinates are not
independent, but one has $\phi^i_{\mu\nu}=\phi^i_{\nu\mu}$ etc.  Local
functions are smooth functions depending on the coordinates $x^\mu$ of
the base space $M$, the fields $\phi^i$, and a finite number of the
jet-coordinates $\phi^i_{\mu_1\dots\mu_k}$. Horizontal forms
involve in addition the differentials $dx^\mu$ which we treat as
anticommuting (Grassmann odd) variables, $dx^\mu dx^\nu=-dx^\nu
dx^\mu$. We also introduce the notation
\begin{eqnarray}
  \label{eq:30}
  (d^{n-p}x)_{\mu_1\dots\mu_p}:=
\frac 1{p!(n-p)!}\, \epsilon_{\mu_1\dots\mu_n}
dx^{\mu_{p+1}}\dots dx^{\mu_n},\quad
\epsilon_{0\dots (n-1)}=1,
\end{eqnarray}
which implies $dx^\alpha (d^{n-p-1}x)_{\mu_1\cdots \mu_{p+1}}  =
 (d^{n-p}x)_{[\mu_1\cdots \mu_{p}}\delta^\alpha_{\mu_{p+1}]}$. 
If the base space is endowed with a metric $g_{\mu\nu}$ (which can
be contained in the set of fields), the Hodge dual
of an horizontal $p$-form $\omega^p$ is defined as $\star\, \omega^p =
\sqrt{|g|}\omega^{\mu_1\dots \mu_p}(d^{n-p}x)_{\mu_1\dots\mu_p}$
where indices are raised with the metric. As a consequence, 
$\star \star\, \omega^p =
(-)^{p(n-p)+s}\omega^p$, where $s$ is the signature of the metric. 

As in \cite{DeDonder1935,Andersonbook}, we define derivatives
$\6^S/\6\phi^i_{\mu_1\dots\mu_k}$ that act on the basic variables
through \bea &&\frac{\6^S
  \phi^j_{\nu_1\dots\nu_k}}{\6\phi^i_{\mu_1\dots\mu_k}} =\delta^j_i\,
\delta^{\mu_1}_{(\nu_1}\dots\delta^{\mu_k}_{\nu_k)}\ ,\quad \frac{\6^S
  \phi^j_{\nu_1\dots\nu_m}}{\6\phi^i_{\mu_1\dots\mu_k}}
=0\quad\mbox{for $m\neq k$},
\nonumber\\[6pt]
&&\frac{\6^S x^\mu}{\6\phi^i_{\mu_1\dots\mu_k}}=0,\quad \frac{\6^S
  dx^\mu}{\6\phi^i_{\mu_1\dots\mu_k}}=0,
\label{derivS}
\eea
where the round parantheses denote symmetrization with weight one,
\[
\delta^{\mu_1}_{(\nu_1}\delta^{\mu_2}_{\nu_2)} =\frac 12
(\delta^{\mu_1}_{\nu_1}\delta^{\mu_2}_{\nu_2}
+\delta^{\mu_1}_{\nu_2}\delta^{\mu_2}_{\nu_1})\quad \mbox{etc.}
\]
For instance, the definition gives explicitly (with $\phi$ any of the
$\phi^i$):
\[
\frac{\6^S\phi_{11}}{\6\phi_{11}}=1\ ,\quad
\frac{\6^S\phi_{12}}{\6\phi_{12}}=
\frac{\6^S\phi_{21}}{\6\phi_{12}}=\frac 12\ ,\quad
\frac{\6^S\phi_{112}}{\6\phi_{112}}=\frac 13\ ,\quad
\frac{\6^S\phi_{123}}{\6\phi_{123}}=\frac 16\ .
\]
We note that the use of these operators takes automatically care of
many combinatorical factors which arise in other conventions, such as
those used in \cite{Olver:1993}.

The vertical differential is defined by \bea d_V=\sum_{k=0}
d_V\phi^i_{\mu_1\dots\mu_k} \frac{\partial^S}{\partial
  \phi^i_{\mu_1\dots\mu_k}}\ ,\label{vertdiff} \eea with Grassmann odd
generators $d_V\phi^i_{\mu_1\dots\mu_k}$, so that $d_V^2=0$.  The
total derivative is the vector field denoted by $\partial_\nu$ and
acts on local functions according to \bea
\partial_\nu=\frac{\partial}{\partial x^\nu}+\sum_{k=0}
\phi^i_{\mu_1\dots\mu_k\nu}\,\frac{\partial^S}{\partial
  \phi^i_{\mu_1\dots\mu_k}}\ . \label{totder} \eea Here $\sum_{k=0}$
means the sum over all $k$, from $k=0$ to infinity, with the summand
for $k=0$ given by $\phi^i_\nu\6/\6\phi^i$, i.e., by definition $k=0$
means ``no indices $\mu_i$''. Furthermore we are using Einstein's
summation convention over repeated indices, i.e., for each $k$ there
is a summation over all tupels $(\mu_1,\dots,\mu_k)$.  Hence, for
$k=2$, the sum over $\mu_1$ and $\mu_2$ contains both the tupel
$(\mu_1,\mu_2)=(1,2)$ and the tupel $(\mu_1,\mu_2)=(2,1)$.  These
conventions extend to all other sums of similar type.

The horizontal differential on horizontal forms is defined by
$d_H=dx^\nu\partial_\nu$. It is extended to the vertical generators in
such a way that $\{d_H,d_V\}=0$.  The derivative of a $n-p$ form $k^{(n-p)} =
k^{[\mu_1\cdots \mu_p]}(d^{n-p}x)_{\mu_1\cdots \mu_p}$ is given by
\[\dH k^{(n-p)} = \d_\rho k^{[\mu_1\cdots
  \mu_{p-1}\rho]}(d^{n-(p-1)}x)_{\mu_1\cdots \mu_{p-1}}.\] 

A vector field of the form $Q^i\partial/\partial\phi^i$, for $Q^i$ a
set of local functions, is called an evolutionary vector field with
characteristic $Q^i$.  Its prolongation which acts on local functions
is \bea \delta_Q =\sum_{k=0}
(\partial_{\mu_1}\dots\partial_{\mu_k}Q^i) \,
\frac{\partial^S}{\partial\phi^i_{\mu_1\dots\mu_k}}\ ,
\label{evvf}
\eea
so that $[\delta_Q,\dH] = 0$.
The Lie bracket of characteristics is defined by
$[Q_1,Q_2]^i=\delta_{Q_1}Q^i_2-\delta_{Q_2}Q^i_1$ and satisfies
\begin{eqnarray}
[\delta_{Q_1},\delta_{Q_2}]=\delta_{[Q_1,Q_2]}.\label{eq:7}
\end{eqnarray}
One has also $[\delta_Q,\dH] = 0$.

An infinitesimal transformation $v$ is defined
by $x^\mu \rightarrow x^\mu+\eps \,c^\mu$ and
$\phi^i \rightarrow \phi^i +\epsilon \, b^i$ with $c^\mu(x,[\phi]), 
b^i(x,[\phi])$ local functions. If we denote by  
$i_c = c^\mu\Q{}{dx^\mu}$ and 
\begin{equation}
\cL_c = i_c \dH + \dH i_c =  c^\mu \d_\mu + \dH c^\mu \Q{}{dx^\mu} 
\label{app:Liec}, 
\end{equation}
the transformation can be extended to act on the horizontal complex as
\[\text{pr}\, v =
\delta_Q  +
\cL_c,\] where $Q^i=b^i-\phi^i_\mu c^\mu$. 
It satisfies $[\text{pr}\,v,\dH]=0$. For example,
a vector field acting on a $n$-form $L \,d^n x$ can be written as
\begin{equation}
\text{pr}\,v\, (L\, d^nx) = \delta_Q L\, d^nx +\dH(c^\mu L
(d^{n-1}x)_\mu).\label{app:deltaL}
\end{equation}

The vector field $\delta_Q$ can be extended so as to also commute with
$d_V$: if we continue to denote the extension by $\delta_{Q}$, the
defining relation
\begin{eqnarray}
  \label{eq:12}
[\delta_Q,\dv]=0
\end{eqnarray}
implies that $\delta_{Q}\dv \phi^i=\dv(Q^i)$. If $i_Q  =
\d_{(\mu)}Q^i \QS{}{\dv\phi^i_{(\mu)}}$, we have 
\begin{eqnarray}
  \{i_{Q},\dv\}=\delta_{Q},\qquad  
[i_{Q_1},\delta_{Q_2}]=i_{[Q_1,Q_2]}.\label{eq:comm2}
\end{eqnarray}
In the context of gauge theories with
$\Phi^\Delta_a=(\phi^i,f^\alpha_a)$ and $\dv$ defined in terms of
$\Phi^\Delta_a$, the relations
\begin{eqnarray}
  \{i_{Q_1},\dv\}=\delta_{Q_1},\ [\delta_{Q_1},\dv]=0,\
  [i_{Q_1},\delta_{Q_2}]=i_{[Q_1,Q_2]},\label{eq:comm}
\end{eqnarray}
continue to hold when $Q_1=R_{f_1},Q_2=R_{f_2}$.

The set of multiindices is simply the set of all tupels
$(\mu_1,\dots,\mu_k)$, including (for $k=0$) the empty tupel. The
tuple with one element is denoted by $\mu_1$ without round
parentheses, while a generic tuple is denoted by $(\mu)$. The length,
i.e., the number of individual indices, of a multiindex $(\mu)$ is
denoted by $|\mu|$.  We use Einstein's summation convention also for
repeated multiindices as in \cite{Andersonbook}. For instance, an
expression of the type $(-\partial)_{(\mu)} K^{(\mu)}$ stands for a
free sum over all tupels $(\mu_1,\dots,\mu_k)$ analogous to the one in
(\ref{totder}),
\[
(-\partial)_{(\mu)} K^{(\mu)}=\sum_{k=0}(-)^k
\partial_{\mu_1}\dots\partial_{\mu_k}K^{\mu_1\dots\mu_k}.
\]
If $Z=Z^{(\mu)}\partial_{(\mu)}$ is a differential operator, its
adjoint is defined by $Z^+=(-\partial)_{(\nu)}[ Z^{(\nu)}\cdot]$ and
its `components' are denoted by $Z^{+(\mu)}$, i.e.,
$Z^+=Z^{+(\mu)}\partial_{(\mu)}$.

More details on the variational bicomplex can be found for instance in
the textbooks
\cite{Andersonbook,Olver:1993,Saunders:1989,Dickey:1991xa}.

\subsection{Higher order Lie-Euler operators}

Except for a different notation, we follow in this and the next
subsection \cite{Andersonbook}.

Multiple integrations by parts can be done using the following.  If
for a given collection $R_i^{(\mu)}$ of local functions, the equality
\begin{eqnarray}
  \label{eq:20}
  \partial_{(\mu)}Q^i P_i^{(\mu)}=\partial_{(\mu)}(Q^i R_i^{(\mu)})
\end{eqnarray}
holds for all local functions $Q^i$, then
\begin{eqnarray}
  \label{eq:21}
  R_i^{(\mu)}=\left(\begin{array}{c}
    |\mu|+|\nu| \\ |\mu| \end{array}\right)(-\partial)_{(\nu)}\,
P_i^{((\mu)(\nu))}\ ,
\label{intparts}
\end{eqnarray}
or, equivalently,
\bea
R_i^{(\mu)}=
\sum_{l=0}\left(\begin{array}{c}
 k+l \\ k \end{array}\right)(-)^l\partial_{\nu_1}\dots\partial_{\nu_l}\,
P_i^{\mu_1\dots\mu_k\nu_1\dots\nu_l}\ ,
\eea
i.e., there is a summation over $(\nu)$
in (\ref{eq:21}) by Einstein's
summation convention for repeated multiindices, and
the multiindex $((\mu)(\nu))$ is the tupel
$(\mu_1,\dots,\mu_k,\nu_1,\dots,\nu_l)$ when
$(\mu)$ and $(\nu)$ are the tupels
$(\mu_1,\dots,\mu_k)$ and $(\nu_1,\dots,\nu_l)$, respectively.
Note that the sum contains only finitely many nonvanishing
terms whenever $f$ is a local function:
if $f$ depends only
on variables with at most $M$ ``derivatives'', i.e., on the
$\phi^i_{(\rho)}$ with $|\rho|\leq M$, the only possibly
nonvanishing summands are those with
$|\nu|\leq M-|\mu|$ ($l\leq M-k$). Conversely, if \eqref{eq:20} holds
for a given collection $R_i^{(\mu)}$ then
\begin{eqnarray}
  \label{eq:22}
  P_i^{(\mu)}=\left(\begin{array}{c}
    |\mu|+|\nu| \\ |\mu| \end{array}\right)\partial_{(\nu)}R_i^{((\mu)(\nu))}
\end{eqnarray}

By definition, when
$P_i^{(\mu)}=\frac{\partial^Sf}{\partial\phi^i_{(\mu)}}$, the higher
order Euler-Lagrange derivatives $\frac{\delta f}{\delta\phi^i_{(\mu)}}$ are
given by the associated $R_i^{(\mu)}$, 
\begin{eqnarray}
\frac{\delta f}{\delta\phi^i_{(\mu)}} =
\left(\begin{array}{c}
    |\mu|+|\nu| \\ |\mu| \end{array}\right)(-\partial)_{(\nu)}\,
\frac{\partial^Sf}{\partial\phi^i_{((\mu)(\nu))}}.\label{higherLie}
\end{eqnarray}
As a consequence, \bea \forall Q^i:\quad
\delta_Q f= \partial_{(\mu)}\Big[ Q^i\,\frac{\delta f}{\delta
  \phi^i_{(\mu)}} \Big]. \label{fund} \eea 
Note also that $\delta/\delta\phi^i$ is
the Euler-Lagrange derivative. The crucial property of these operators
is that they ``absorb total derivatives'', \bea
&|\mu|=0:& \frac{\delta (\partial_\nu f)}{\delta \phi^i}=0,\label{eA7}\\
&|\mu|>0:& \frac{\delta (\partial_\nu f)}{\delta \phi^i_{(\mu)}}
=\delta^{(\mu}_\nu\frac{\delta f}{\delta \phi^i_{(\mu^\prime))}}\
,\quad (\mu)=(\mu(\mu^\prime)),\label{eA8} \eea where, e.g.,
\[
\delta^{(\mu}_\nu\frac{\delta f}{\delta
\phi^i_{\lambda)}}=\frac{1}{2}\big(\delta^{\mu}_\nu\frac{\delta f}{\delta
\phi^i_{\lambda}}+\delta^{\lambda}_\nu\frac{\delta f}{\delta
\phi^i_{\mu}}\big).
\]
It may be also deduced that
\begin{eqnarray}
\varQ{(\d_\nu f)}{\phi^i_{\rho(\mu)}} =
\frac{1}{|\mu|+1}\delta^\rho_\nu
\varQ{f}{\phi^i_{(\mu)}}+\frac{|\mu|}{|\mu|+1}\delta^{(\mu_1}_\nu
\varQ{f}{\phi^i_{\rho \mu_2 \cdots \mu_{|\mu|}) } }.
\label{higherLieprop}
\end{eqnarray}

By considering the particular case
where \eqref{eq:20}, \eqref{eq:21} are used in terms
of $Q_2$ with
\begin{eqnarray}
P_i^{(\mu)}[\vddl{\omega^n}{\phi}]=\ddl{^S
  Q^j_1}{\phi^i_{(\mu)}}\vddl{\omega^n}{\phi^j},\label{eq:25}
\end{eqnarray}
we get
$\delta_{Q_2}(Q^j_1)\vddl{\omega^n}{\phi^j}=
\partial_{(\mu)}\Big(Q_2^i R^{(\mu)}_i[\vddl{\omega^n}{\phi}]\Big)$.
Splitting the term without derivatives on the RHS from the others
and defining
\begin{eqnarray}
 \hspace*{-1cm} T_{Q_1}[Q_2,\vddl{\omega^n}{\phi}]&=&\partial_{(\mu)}\Big(Q_2^i
  R^{(\mu)\nu}_i[\frac{\partial}{\partial
    dx^\nu}\vddl{\omega^n}{\phi}]\Big),\nonumber
 \\ &=&\left(\begin{array}{c}
    |\mu|+1+|\rho| \\ |\mu|+1 \end{array}\right)
\partial_{(\mu)}\Bigg(Q_2^i (-\6)_{(\rho)}
\Big(\ddl{^SQ^j_1}{\phi^i_{((\mu)(\rho)\nu)}}\frac{\partial}{\partial
    dx^\nu}\vddl{\omega^n}{\phi^j}\Big)
\Bigg),\label{eq:27}
\end{eqnarray}
gives
\begin{eqnarray}
\delta_{Q_2}(Q^j_1)\vddl{\omega^n}{\phi^j}=Q^i_2 R_i+\dH
T_{Q_1}[Q_2,\vddl{\omega^n}{\phi}], \quad R_i=(-\6)_{(\nu)}\Big(
\ddl{^SQ_1^j}{\phi^i_{(\nu)}}\vddl{\omega^n}{\phi^j}\Big).\label{eq:28}
\end{eqnarray}

We also need the definition
\begin{multline}
  \label{eq:34a}
  \delta_{Q_3}T_{Q_1}[Q_2,\vddl{\omega^n}{\phi}]=T_{Q_1}[\delta_{Q_3}Q_2,
\vddl{\omega^n}{\phi}]+
T_{Q_1}[Q_2,\delta_{Q_3}\vddl{\omega^n}{\phi}]+
\\+T_{\delta_{Q_3}Q_1}[Q_2,\vddl{\omega^n}{\phi}]-Y_{Q_1,Q_3}[Q_2,
\vddl{\omega^n}{\phi}],
\end{multline}
where
\begin{multline}
  \label{eq:16bis}
Y_{Q_1,Q_3}[Q_2,\vddl{\omega^n}{\phi}]=
\left(\begin{array}{c}|\mu|+|\rho|+1\\ |\mu|+1\end{array}\right)
\6_{(\mu)}
\Big(Q_2^i(-\6)_{(\rho)}\\\big(\frac{\partial}{\partial
    dx^\nu}\vddl{\omega^n}{\phi^j}\frac{\6^S \partial_{(\sigma)} Q^k_3}
{\6\phi^i_{((\mu)(\rho)\nu)}}\frac{\6^S Q_1^j}
{\6\phi^k_{(\sigma)}}\big)
\Big).
\end{multline}

\subsection{Horizontal homotopy operators}

Define
\bea
I^{p}_{\dv\phi}\omega^{p,s}=
\frac{|\mu|+1}{n-p+|\mu|+1}\ \partial_{(\mu)}\Big(
\dv\phi^i
\frac{\delta}{\delta\phi^i_{((\mu)\nu)}}\,\frac{\partial
\omega^{p,s}}{\partial dx^\nu}\Big)
\label{phihomotopy}
\eea for $\omega^{p,s}$ a $p,s$-form. Note that there is a
summation over $(\mu)$ by Einstein's summation convention.  The
following result (see e.g.~\cite{Andersonbook}) is the key for showing
local exactness of the horizontal part of the variational bicomplex:
\bea &0\leq p< n:&\delta_Q \omega^{p,s}=
I^{p+1}_{Q}(d_H\omega^{p,s})+d_H(I^p_{Q}\omega^{p,s}),
\label{cc1}\\
&p=n:& \delta_Q \omega^{n,s} =Q^i \frac{\delta\omega^{n,s}}{\delta\phi^i}
+d_H(I^n_{Q}\omega^{n,s}).\label{cc2} \eea
This last relation is sometimes referred to as ``the first variational
formula''. 
Similarily,
\bea &0\leq p< n:&\dv \omega^{p,s}=
I^{p+1}_{\dv\phi}(d_H\omega^{p,s})-d_H(I^p_{\dv\phi}\omega^{p,s});
\label{cc1a}\\
&p=n:& \dv \omega^{n,s} =\dv\phi^i \frac{\delta\omega^{n,s}}{\delta\phi^i}
-d_H(I^n_{\dv\phi}\omega^{n,s}).\label{cc2a} \eea

In the context of the extended jet-bundle of gauge theories, we will
also use the following homotopy operators that only involve the gauge
parameters: for local functions $G^\alpha$
\begin{eqnarray}
  \label{eq:45}
I^{p}_{G}\omega^{p,s}_f= \frac{|\lambda|+1}{n-p+|\lambda|+1}
\partial_{(\lambda)}  \Big(G^\alpha
  \vddl{}{f^\alpha_{,(\lambda)\rho}} \ddl{\omega^{p,s}_f}{dx^\rho}\Big).
\end{eqnarray}
When applied to $p,s$ forms that are linear and homogeneous in
$f^\alpha$ and its derivatives, we have
\begin{eqnarray}
  \label{eq:2}
 &0\leq p< n:& \omega^{p,s}_G=I^{p+1}_{G}(\dH\omega^{p,s}_f)+\dH
 (I^{p}_{G}\omega^{p,s}_f), \\
&p=n:& \omega^{n,s}_G=G^\alpha\vddl{\omega^{n,s}_f}{f^\alpha}+\dH
 (I^{n}_{G}\omega^{n,s}_f).
\end{eqnarray}

\subsection{Commutation relations}
\label{sec:comm-relat}

Starting from $\delta_{Q_1}\delta_{Q_2}\omega^n-\delta_{Q_2}\delta_{Q_1}\omega^n=
\delta_{[Q_1,Q_2]}\omega^n$ and using \eqref{cc2} both on the inner terms
of the LHS and on the RHS gives
\begin{eqnarray}
  \label{eq:50}
  Q_2^i\delta_{Q_1}\vddl{\omega^n}{\phi^i}-Q_1^i\delta_{Q_2}
\vddl{\omega^n}{\phi^i} = \dH( I^n_{[Q_1,Q_2]}\omega^n-
\delta_{Q_1}I^n_{Q_2}\omega^n+\delta_{Q_2}I^n_{Q_1}\omega^n ).
\end{eqnarray}

Using $\dv(
\delta_Q\omega)=\delta_Q(\dv\omega)$, we get
$\partial_{(\mu)}(\dv
\phi^i \vddl{\delta_Q\omega}{\phi^i_{(\mu)}})
=\partial_{(\mu)}(\delta_Q(\dv
\phi^i\vddl{\omega}{\phi^i_{(\mu)}}))$, which can be written as
$\partial_{(\mu)}(\dv \phi^i
[\vddl{}{\phi^i_{(\mu)}},\delta_Q]{\omega})
=\partial_{(\mu)}(\dv Q^i\vddl{\omega}{\phi^i_{(\mu)}})$.
Applying $\vddl{}{\dv\phi^i_{\mu_1\dots\mu_k}}$ gives
\begin{eqnarray}
    \label{eq:1}
[\vdl{\phi^i_{\mu_1\dots\mu_k}},\delta_Q]\omega&=&\sum_{l\leq k}
\left(\begin{array}{c}
    l+|\nu| \\ l \end{array}\right)(-\partial)_{(\nu)}\Big(
\frac{\6^S
 Q^j}{\6\phi^i_{((\nu)\mu_1\dots\mu_l}}
\vddl{\omega}{\phi^j_{\mu_{l+1}\dots\mu_k)}}
\Big).
\end{eqnarray}
In particular,
\begin{eqnarray}
  \label{eq:3}
  [\vdl{\phi^i},\delta_Q]\omega=(-\partial)_{(\nu)}\Big(
\frac{\6^S
 Q^j}{\6\phi^i_{(\nu)}}
\vddl{\omega}{\phi^j}
\Big).
\end{eqnarray}
When combined with \eqref{eq:28}, we get
\begin{eqnarray}
  \label{eq:32}
  Q^i_2[\delta_{Q_1},\vdl{\phi^i}]\omega^n=
-\delta_{Q_2}Q^j_1\vddl{\omega^n}{\phi^j} +\dH
T_{Q_1}[Q_2,\vddl{\omega^n}{\phi}].
\end{eqnarray}

Similarly, applying $\vddl{}{\dv\phi^i_{\mu_1\dots\mu_k}}$ to
$\partial_{(\mu)}(\dv \phi^i
\vddl{(\delta_Q\omega)}{\phi^i_{(\mu)}})=\partial_{(\mu)}(\dv
(Q^i\vddl{\omega}{\phi^i_{(\mu)}}))$, gives
\begin{eqnarray}
\vdl{\phi^i_{\mu_1\dots\mu_k}}(\delta_Q\omega)
&=&\sum_{l\leq
  k}\vdl{\phi^i_{(\mu_1\dots\mu_l}}\Big(Q^j\vddl{\omega}{\phi^j_{\mu_{l+1}\dots
  \mu_k)}}\Big).
\end{eqnarray}
Applying $\vddl{}{\dv\phi^i_{(\lambda)}}$ to
$\dv\vddl{\omega^n}{\phi^j}=\vddl{}{\phi^j}(\dv\phi^i\vddl{\omega^n}{\phi^i})$,
we also get
\begin{eqnarray}
  \label{eq:39}
  \vddl{}{\phi^i_{(\lambda)}}\vddl{\omega^n}{\phi^j}=(-)^{|\lambda|}\frac{\6^S}
{\6\phi^j_{(\lambda)}}\vddl{\omega^n}{\phi^i}.
\end{eqnarray}

Starting from $\dH ([\delta_{Q_1},I^n_{Q_2}]\omega^n)=
\delta_{[Q_1,Q_2]}\omega^n-\delta_{Q_1}Q^i_2\vddl{\omega^n}{\phi^i}-
Q^i_2\delta_{Q_1}\vddl{\omega^n}{\phi^i}+
Q^i_2\vddl{(\delta_{Q_1}\omega^n)}{\phi^i}$ and using
\eqref{eq:32} to compute the last two terms, we find
\begin{eqnarray}
  \label{eq:19}
  \dH ([\delta_{Q_1},I^n_{Q_2}]\omega^n)=\dH
  (I^n_{[Q_1,Q_2]}\omega^n)-\dH T_{Q_1}[Q_2,\vddl{\omega^n}{\phi}].
\end{eqnarray}
Similarly, for $p<n$, by evaluating $\dH
([\delta_{Q_1},I^p_{Q_2}]\omega^p)$ one finds
\begin{eqnarray}
  \label{eq:24}
  \dH
([\delta_{Q_1},I^p_{Q_2}]\omega^p)=\dH (I^p_{[Q_1,Q_2]}\omega^p)+
(I^{p+1}_{[Q_1,Q_2]}-[\delta_{Q_1},I^{p+1}_{Q_2}])(\dH\omega^p).
\end{eqnarray}
By the same type of arguments, one shows
\begin{multline}
  \label{eq:29}
  \dH\Big(
\delta_{Q_1}(I^n_{Q_2}\omega^n)-(1\leftrightarrow
2)\Big)=\\= \dH\Big(I^n_{[Q_1,Q_2]}\omega^n-I^n_{Q_1}(\delta_{Q_2}\omega^n)-
  T_{Q_1}[Q_2,\vddl{\omega^n}{\phi}]-(1\leftrightarrow 2)\Big),
\end{multline}
\begin{multline}
  \label{eq:29bis}
  \dH\Big(
\delta_{Q_1}(I^p_{Q_2}\omega^p)-(1\leftrightarrow
2)\Big)=\\= \dH\Big(I^p_{[Q_1,Q_2]}\omega^p\Big)+(I^{p+1}_{[Q_1,Q_2]}
-\delta_{Q_1}I^{p+1}_{Q_2}+\delta_{Q_2}I^{p+1}_{Q_1})(\dH \omega^p).
\end{multline}

\subsection{Properties of the invariant presymplectic $(n-1,2)$ form}
\label{sec:prop-w_vddl-1}

Let us define the $(n-1,2)$-forms 
\begin{eqnarray}
  \label{eq:66a}
W_{{\delta\omega^n}/{\delta\phi}}=-
\half I^n_{\dv\phi}\big(\dv\phi^i
  \vddl{\omega^n}{\phi^i}\big),  \qquad
\Omega_{\omega^n}=\dv I^n_{\dv\phi}\omega^n,
\end{eqnarray}
and the $(n-2,2)$-form 
\begin{eqnarray}
E_{\omega^n}=\half
I^{n-1}_{\dv\phi}I^n_{\dv\phi}\omega^n\label{eq:26}.
\end{eqnarray}

Using $[\dv,I^n_{\dv\phi}]=0$, \eqref{cc1a} and \eqref{cc2a}
imply
\begin{eqnarray}
  \label{eq:80}
  \half I^n_{\dv\phi}\big(\dv\phi^i
  \vddl{\omega^n}{\phi^i}\big)=\dv I^n_{\dv\phi}\omega^n +\half \dH
  (I^{n-1}_{\dv\phi}I^n_{\dv\phi}\omega^n), 
\end{eqnarray}
so that
\begin{eqnarray}
  \label{eq:79}
  -W_{{\delta\omega^n}/{\delta\phi}}=\Omega_{\omega^n}+\dH
E_{\omega^n},\qquad\dv \Omega_{\omega^n}=0.
\end{eqnarray}
$\Omega_{\omega^n}$ is the
presymplectic $(n-1,2)$ form usually used in the context of covariant phase
space methods. The ``second variational formula'', obtained by
applying $\dv$ to \eqref{cc2a}, can be combined with
\eqref{eq:79} to give 
\begin{eqnarray}
  \label{eq:13}
  \dv\phi^i\dv \vddl{\omega^n}{\phi^i}=\dH \Omega_{\omega^n}=-\dH
  W_{{\delta\omega^n}/{\delta\phi}}. 
\end{eqnarray}
Our surface charges are related to $W_{{\delta\omega^n}/{\delta\phi}}$, which
is $\dv$-closed only up to a $\dH$ exact term, 
\begin{eqnarray}
\dv W_{{\delta\omega^n}/{\delta\phi}}=\dH\dv E_{\omega^n}.\label{eq:57}
\end{eqnarray}
Contrary to $\Omega_{\omega^n}$, it involves the Euler-Lagrange
derivatives and is thus independent of $\dH$ exact $n$-forms that are
added to $\omega^n$. For this reason, we call
$W_{{\delta\omega^n}/{\delta\phi}}$ the invariant presymplectic
$(n-1,2)$ form.

When $\omega^n = \dH \omega^{n-1}$, we have
\begin{equation}
E_{\dH \omega^{n-1}} = \dv I^{n-1}_{\dv\phi}\omega^{n-1}+\dH
I_{\dv\phi}^{n-2}I_{\dv\phi}^{n-1}\omega^{n-1}.
\end{equation}

The following proposition, proven at the end of this section, is
crucial and generalizes corresponding  
results in \cite{Barnich:2001jy,Julia:2002df}:
\begin{prop}\label{la}
\begin{eqnarray}
  \label{eq:4}
\hspace*{-.3cm} W_{\vddl{\omega^n}{\phi}}[Q_1,Q_2]=
\left(\begin{array}{c}|\mu|+|\rho|+1\\ |\mu|+1\end{array}\right)
\6_{(\mu)}
\Big(Q_1^i(-\6)_{(\rho)}(Q^j_2\frac{\6^S }
{\6\phi^i_{((\mu)(\rho)\nu)}}\frac{\partial}{\partial
    dx^\nu}\vddl{\omega^n}{\phi^j} )
\Big).
\end{eqnarray}
\end{prop}
Together with the definition of the homotopy operator
\eqref{phihomotopy}, the definition of the higher order Euler-Lagrange
derivatives and definition \eqref{eq:27}, we find from~\eqref{eq:4}
that
\begin{eqnarray}
  \label{eq:36}
  I^n_{Q_1}(Q_2^i\vddl{\omega^n}{\phi^i})=
W_{{\delta\omega^n}/{\delta\phi}}[Q_1,Q_2]+T_{Q_2}[Q_1,\vddl{\omega^n}{\phi}].
\end{eqnarray}
We then can define 
\begin{multline}
  \label{eq:34bis}
  \delta_{Q_3}W_{{\delta\omega^n}/{\delta\phi}}[Q_1,Q_2]=
W_{{\delta\omega^n}/{\delta\phi}}[\delta_{Q_3}Q_1,Q_2]+
W_{{\delta\omega^n}/{\delta\phi}}[Q_1,\delta_{Q_3}Q_2]+\\
+ Z_{{\delta\omega^n}/{\delta\phi}}[Q_1,Q_2,Q_3]
\end{multline}
where
\begin{multline}
  \label{eq:16ter}
Z_{{\delta\omega^n}/{\delta\phi}}[Q_1,Q_2,Q_3]=
\left(\begin{array}{c}|\mu|+|\rho|+1\\ |\mu|+1\end{array}\right)
\6_{(\mu)} \Big(Q_1^i(-\6)_{(\rho)}\\\big(Q^j_2\partial_{(\sigma)}
Q^k_3 \frac{\6^S } {\6\phi^k_{(\sigma)}}\frac{\6^S }
{\6\phi^i_{((\mu)(\rho)\nu)}}\frac{\partial}{\partial
    dx^\nu}\vddl{\omega^n}{\phi^j}\big)
\Big).
\end{multline}
Note also that it follows from proposition \bref{la} that, in
the case of first order theories, $W_{{\delta\cL}/{\delta\phi}}$ coincides
with the covariant symplectic density $\hat \omega$ considered in
\cite{Julia:2002df},
\begin{equation}
W_{{\delta\omega^n}/{\delta\phi}} = \half \dv\phi^i 
\dv\phi^j \QS{}{\phi^i_\nu}\left( \Q{}{dx^\nu}
\varQ{\omega^n}{\phi^j}\right).
\end{equation}

Additional relations are obtained by applying $\dH$ to \eqref{eq:36} and
using \eqref{eq:32}, which gives
\begin{multline}
  \dH\Big(W_{{\delta\omega^n}/{\delta\phi}}[Q_1,Q_2]+
T_{Q_2}[Q_1,\vddl{\omega^n}{\phi}]-
T_{Q_1}[Q_2,\vddl{\omega^n}{\phi}]\Big)=\\=\dH
\Big(-I^n_{[Q_1,Q_2]}\omega^n+I^n_{Q_1}
(\delta_{Q_2}\omega^n)-I^n_{Q_2}
(\delta_{Q_1}\omega^n)\Big).  \label{eq:63a}
\end{multline}
Starting from
$[\delta_{Q_1},\delta_{Q_2}]\omega^n= \delta_{[Q_1,Q_2]}\omega^n$ and
using \eqref{cc2} on the outer terms of the LHS, 
\eqref{eq:36} and \eqref{eq:32}
gives
\begin{multline}
  \label{eq:65a}
  Q_1^i\delta_{Q_2}\vddl{\omega^n}{\phi^i}-Q_2^i\delta_{Q_1}
\vddl{\omega^n}{\phi^i} = \dH( I^n_{[Q_1,Q_2]}\omega^n-2
W_{{\delta\omega^n}/{\delta\phi}}[Q_1,Q_2]-\\-
\delta_{Q_1}I^n_{Q_2}\omega^n+\delta_{Q_2}I^n_{Q_1}\omega^n ).
\end{multline}
Adding to \eqref{eq:50} gives in particular
\begin{eqnarray}
  \label{eq:65b}
  \dH W_{{\delta\omega^n}/{\delta\phi}}[Q_1,Q_2]=\dH( I^n_{[Q_1,Q_2]}\omega^n-
\delta_{Q_1}I^n_{Q_2}\omega^n+\delta_{Q_2}I^n_{Q_1}\omega^n ).
\end{eqnarray}

\vspace{1cm}

{\bf Proof of proposition \bref{la}:} 

Let $R[Q_1,Q_2]$ be the RHS of~\eqref{eq:4}. By applying
$i_{Q_2}i_{Q_1}$ to $W_{{\delta\omega^n}/{\delta\phi}}$, it
follows that proposition \bref{la} amounts to showing
\begin{eqnarray}
 \label{eq:67}
 R[Q_1,Q_2]=-R[Q_2,Q_1].
\end{eqnarray}
Splitting the derivatives $(\mu)$ in those acting on $Q^i_1$,
denoted by $(\alpha)$, and in those acting on the remaining
expression, denoted by $(\mu^\prime)$ and regrouping the indices
$((\mu^\prime)(\rho)) \equiv (\sigma)$, we get,
\begin{multline}
R[Q_1,Q_2]=\sum_{|\alpha|\geq 0}\sum_{|\sigma| \geq
|\mu^\prime|\geq
 0}\left(\begin{array}{c}|\sigma|+|\alpha|+1\\
|\mu^\prime|+|\alpha|+1\end{array}\right)
\left(\begin{array}{c}|\mu^\prime|+|\alpha|\\
|\alpha|\end{array}\right)(-)^{|\mu^\prime|}\\
\d_{(\alpha)}Q^i_1(-\d)_{(\sigma)}
\Big(Q_2^j\frac{\partial^S}{\d\phi^i_{((\sigma)(\alpha)\nu)}}
\frac{\partial}{\partial  dx^\nu}\vddl{\omega^n}{\phi^j}\Big).
\end{multline}
We now evaluate $\sum_{|\sigma| \geq |\mu^\prime|\geq 0}$ as
$\sum_{|\sigma| \geq 0}\sum_{|\mu^\prime|=0}^{|\sigma|}$ and use
the fact that
\begin{eqnarray}
 \label{eq:33}
 \sum_{|\mu^\prime|=0}^{|\sigma|} \left(\begin{array}{c}|\sigma|+|\alpha|+1\\
|\mu^\prime|+|\alpha|+1\end{array}\right)
\left(\begin{array}{c}|\mu^\prime|+|\alpha|\\
|\alpha|\end{array}\right)(-)^{|\mu^\prime|}=1,
\end{eqnarray}
for all $|\alpha|,|\sigma|$, so that
\begin{equation}
R[Q_1,Q_2] =\6_{(\alpha)}Q_1^i
(-\6)_{(\sigma)}\Big(Q^j_2\frac{\6^S}
{\6\phi^i_{((\alpha)(\sigma)\nu)}} \frac{\partial}{\partial
dx^\nu}\vddl{\omega^n}{\phi^j} \Big).\label{eq:app1}
\end{equation}
Expanding the $\sigma$ derivatives,
\begin{equation}
R[Q_1,Q_2] =
\d_{(\alpha)}Q^i_1\; \d_{(\beta)}Q^j_2 \; C_{ij}^{(\alpha)(\beta)},
\end{equation}
where
\begin{equation}
C_{ij}^{(\alpha)(\beta)} = (-)^{|\beta|} \left(\begin{array}{c}|\rho|+|\beta|\\
|\beta|\end{array}\right)(-\d)_{(\rho)}
\frac{\d^S}{\d\phi^i_{(\alpha)(\beta)(\rho)\nu}}\varQ{}{\phi^j}\frac{\d}{\d
dx^\nu }\omega^n.\label{defCij}
\end{equation}
Antisymmetry~\eqref{eq:67} amounts to prove that
\begin{equation}
C_{ij}^{(\alpha)(\beta)} =
-C_{ji}^{(\beta)(\alpha)}.\label{eq:antisym}
\end{equation}
From equation \eqref{eq:39}, we get
\begin{eqnarray}
 \label{eq:75}
&&  C_{ij}^{(\alpha)(\beta)} = -(-)^{|\alpha|} 
\left(\begin{array}{c}|\rho|+|\beta|\\
|\beta|\end{array}\right)\d_{(\rho)}
\vddl{}{\phi^i_{(\alpha)(\beta)(\rho)\nu}}\varQ{}{\phi^i}\frac{\d}{\d
dx^\nu }\omega^n.
\end{eqnarray}
Using the definition of higher order Lie
operators~\eqref{higherLie} we get
\begin{eqnarray}
C_{ij}^{(\alpha)(\beta)} &=& - \sum_{|\sigma^\prime| \geq |\rho| \geq 0} (-)^{|\alpha|+|\sigma^\prime|+|\rho|} \left(\begin{array}{c}|\rho|+|\beta|\\
|\beta|\end{array}\right) \nonumber \\
&&\left(\begin{array}{c}|\alpha|+|\beta|+|\sigma^\prime|+1\\
|\alpha|+|\beta|+|\rho|+1\end{array}\right) \d_{(\sigma^\prime)}
\frac{\d^S}{\phi^j_{(\alpha)(\beta)(\sigma^\prime)\nu}}\varQ{}{\phi^i}\frac{\d}{\d
dx^\nu }\omega^n.
\end{eqnarray}
Evaluating $\sum_{|\sigma^\prime| \geq |\rho| \geq 0}$ as
$\sum_{|\sigma^\prime| \geq 0}\sum_{|\rho|=0}^{|\sigma^\prime|}$
and using the equality
\begin{eqnarray}
 \label{eq:81}
 \sum_{|\rho|=0}^{|\sigma^\prime|}(-)^{|\rho|}\left(\begin{array}{c}|\rho|+|\beta|\\
|\beta|\end{array}\right)\left(\begin{array}{c}|\alpha|+
|\beta|+|\sigma^\prime|+1\\ 
|\alpha|+|\beta|+|\rho|+1\end{array}\right)=
\left(\begin{array}{c}|\sigma^\prime|+|\alpha|\\
|\alpha|\end{array}\right),
\end{eqnarray}
we finally obtain
\begin{eqnarray}
C_{ij}^{(\alpha)(\beta)}\hspace{-6pt} &=&\hspace{-6pt} -(-)^{|\alpha|}
\left(\hspace{-3pt}\begin{array}{c}|\sigma^\prime|+|\alpha|\\
|\alpha|\end{array}\hspace{-3pt}\right)(-\d)_{(\sigma^\prime)}
\frac{\d^S}{\d\phi^j_{(\alpha)(\beta)(\sigma^\prime)\nu}}\varQ{}{\phi^i}\frac{\d}{\d
dx^\nu }\omega^n.
\end{eqnarray}
Comparing with~\eqref{defCij}, we have~\eqref{eq:antisym} as it
should. \qed

\section{Algebra of surface charge 1-forms}
\label{appb}

For compactness, let us define a generalized gauge transformation through
$$\delta^T_{f_1}\Phi^\Delta_2=(R^i_{f_1},[f_1,f_2]^\alpha).$$
According to the same reasoning that led to \eqref{eq:5}, combined
with \eqref{1.18} and the definition \eqref{sec1cur} of Noether
currents for gauge symmetries, we get
\begin{eqnarray}
  \label{eq:9}
  d_H\Big(\delta^T_{f_1}S_{f_2}
  -M_{f_1,f_2}-T_{R_{f_1}}[R_{f_2},\vddl{\cL}{\phi}]\Big) =0.
\end{eqnarray}
Applying the contracting homotopy with respect to the gauge parameters
$f^\alpha_1$ now gives
\begin{eqnarray}
  \label{eq:10}
\delta^T_{f_1}S_{f_2}
=M_{f_1,f_2}+T_{R_{f_1}}[R_{f_2},\vddl{\cL}{\phi}]+
d_HN_{f_1,f_2},
\end{eqnarray}
where
\begin{eqnarray}
  \label{eq:11}
  N_{f_1,f_2}[\vddl{\cL}{\phi}]=I^{n-1}_{f_1}\big(
\delta^T_{f_1}S_{f_2}
-M_{f_1,f_2}-T_{R_{f_1}}[R_{f_2},\vddl{\cL}{\phi}]\big).
\end{eqnarray}
By applying $1=\{I_{f_1},\dH\}$ to $\delta^T_{f_1} k_{f_2}$ and using
$\dH k_{f_2}= -\dv^\phi S_f+ I^{n}_{\dv\phi}(\dH S_f)$,
we get
\begin{eqnarray}
  \label{eq:13b}
\delta^T_{f_1}  k_{f_2}
  = I^{n-1}_{f_1}\Big(-\delta^T_{f_1} \dv^\phi
 S_{f_2}+\delta^T_{f_1}( I^n_{\dv\phi}
 (\dH S_{f_2}))\Big)
+\dH (\cdot).
\end{eqnarray}
Using the properties of the homotopy operators, the expression inside
the parenthesis of RHS of \eqref{eq:13b} becomes
\begin{multline}
  \label{eq:100}
-\delta^T_{f_1} \dv^\phi  S_{f_2}+\delta^T_{f_1}( I^n_{\dv\phi}
 (\dH S_{f_2})) = -[\delta^T_{f_1}, \dv^\phi]  S_{f_2}+
[\delta^T_{f_1}, I^n_{\dv\phi}](\dH S_{f_2})
+\\ +\dH I^{n-1}_{\dv \phi}(\delta^T_{f_1}S_{f_2}).
\end{multline}
{}From equation \eqref{eq:10}, we get
\begin{multline}
  \label{eq:15}
\delta^T_{f_1}  k_{f_2}=I^{n-1}_{f_1}(-\delta_{R_{\dv f_1}}S_{f_2}+
 [\delta^T_{f_1},
I^n_{\dv\phi}](\dH S_{f_2}))+\dv^\phi N_{f_1,f_2}+\\+
I^{n-1}_{\dv\phi}(M_{f_1,f_2}+
  T_{R_{f_1}}[R_{f_2},\vddl{\cL}{\phi}])
+ \dH (\cdot).
\end{multline}
Using \eqref{eq:40}, \eqref{eq:34a}, and \eqref{eq:34bis} for the
direct computation of $[\delta^T_{f_1}, I^n_{\dv\phi}](\dH S_{f_2})$
gives
 \begin{multline}
   \label{eq:38}
   [\delta^T_{f_1}, I^n_{\dv\phi}](\dH S_{f_2})=
W_{{\delta\cL}/{\delta\phi}}[R_{f_2},\dv^\phi
R_{f_1}]+T_{R_{f_2}}[\dv^\phi
R_{f_1},\vddl{\cL}{\phi}]-\\-
Y_{R_{f_2},R_{f_1}}[\dv\phi,\vddl{\cL}{\phi}]
+Z_{{\delta\cL}/{\delta\phi}}[R_{f_2},\dv\phi,R_{f_1}] -
W_{\delta_{R_{f_1}}\varQ{\cL}{\phi}}[R_{f_2},\dv\phi].
 \end{multline}
If
\begin{multline}
\cT_{f_1,f_2}[\dv\phi]:=
 \Bigg[
I^{n-1}_{f_1}\Bigg( -\delta_{R_{\dv f_1}}S_{f_2}
+W_{{\delta\cL}/{\delta\phi}}[\dv^\phi
R_{f_1},R_{f_2}]+T_{R_{f_2}}[\dv^\phi
R_{f_1},\vddl{\cL}{\phi}]-\\-Y_{R_{f_2},R_{f_1}}
[\dv\phi,\vddl{\cL}{\phi}]
+Z_{{\delta\cL}/{\delta\phi}}[\dv\phi,R_{f_2},R_{f_1}] -
W_{\delta_{R_{f_1}}\varQ{\cL}{\phi}}[\dv\phi,R_{f_2}] \Bigg) +\\
+\dv^\phi N_{f_1,f_2}+ I^{n-1}_{\dv\phi}(M_{f_1,f_2}+
  T_{R_{f_1}}[R_{f_2},\vddl{\cL}{\phi}]) \Bigg],\label{c2}
\end{multline}
we finally have
\begin{eqnarray}
 \delta_{R_{f_1}} k_{f_2}[\dv \phi] = -k_{[f_1,f_2]}[\dv \phi]
+ \cT_{f_1,f_2}[\dv\phi]
+d_H(\cdot).\label{main1}
\end{eqnarray}
Note that $\cT_{f_1,f_2}[\dv\phi]=0$ if (i) $\phi^s$ is a solution to
the Euler-Lagrange equations of motion, (ii) $R_{f_2}|_{\phi^s}=0$
and, (iii) $\dv\phi$ is tangent to the space of solutions at
$\phi^s$. This proves Proposition \bref{lem11}. \qed

\section{Integrability implies algebra}
\label{appc}

Owing to \eqref{4.13}
\begin{eqnarray}
  \label{eq:83a}
  \dv^s k_{f}[\dv^s\phi]\approx 
I_{\dv f}^{n-1} (W[\dv^s \phi,R_f])
- I_{f}^{n-1} (\dv^{s}W[\dv^s \phi,R_f])+\dH(\cdot).
\end{eqnarray}
After application of $i_{R_{g_1}}$ we find that the integrability
condition \eqref{eq:92a} for the charge associated to $g_2$ reads
\begin{multline}
  \label{eq:63}
  \oint_{S^{r,t}} 
\Big(I_{\dv g_2}^{n-1} (W[R_{g_2},R_{g_1}])+
I_{g_2}^{n-1}(\delta_{R_{g_1}}W[R_{g_2},\dv^s
\phi]-\\-
\dv^{s} (W[R_{g_2},R_{g_1}]))\Big)\approx o(r^0).
\end{multline}
Notice that it follows from \eqref{eq:29abis} that 
\begin{eqnarray}
  \label{eq:86}
k_{[f_1,f_2]}[\dv^s\phi]\approx
I_{f_2}^{n-1}(W[\dv^s\phi,R_{[f_1,f_2]}])
+\dH(\cdot)\label{eq:69c},
\end{eqnarray}
and 
\begin{multline}
I_{f_1}^{n-1}(\delta_{R_{f_2}}
W[\dv\phi^s,R_{f_1}])\approx I_{f_2}^{n-1}(\delta_{R_{f_2}}
W[\dv\phi^s,R_{f_1}])-\\-
I_{\dv f_2}^{n-1}(W[R_{f_2},R_{f_1}])+\dH(\cdot).\label{eq:68}
\end{multline}
Applying $I^{n-1}_{f_2}$ to  \eqref{eq:57} contracted with 
$R_{f_1},R_{f_2}$ and using \eqref{eq:68}, we get
\begin{multline}
  \label{eq:84}
  I^{n-1}_{f_2}\Big(\dv^{s} W[R_{f_2},R_{f_1}]-
\delta_{R_{f_1}}W[R_{f_2},\dv^s\phi]-
W[R_{[f_1,f_2]},\dv^s\phi]\Big)\\ +
I^{n-1}_{f_1}(
\delta_{R_{f_2}}W[R_{f_1},\dv^s\phi]) +I_{\dv f_2}^{n-1}(W[R_{f_1},R_{f_2}])\approx 
i_{R_{f_1}}i_{R_{f_2}}\dv^s  E+\dH(\cdot).
\end{multline}
Integrating over ${S^{r,t}}$ and using gauge parameters $g_1,g_2$ so
that \eqref{eq:63} can be used to treat the second and fourth terms on
the RHS, we find
\begin{multline}
  \label{eq:61}
  \oint_{S^{r,t}} \Big(- I^{n-1}_{\dv g_1}(W[R_{g_1},R_{g_2}])+I^{n-1}_{
    g_1}(\dv^{s}W[R_{g_1},R_{g_2}])\Big)\approx\\\approx\oint_{S^{r,t}}
  \Big(I^{n-1}_{g_2}W[R_{[g_1,g_2]},\dv^s\phi]\Big) + o(r^0),
\end{multline}
if and only if 
$\oint_{S^{r,t}} i_{R_{g_2}}i_{R_{g_1}}\dv^s  E_{\cL}\approx o(r^0)$,
which holds as a consequence of assumption \eqref{eq:109}. 
Using \eqref{4.13} on the LHS and \eqref{eq:69c} on the RHS, we
get \eqref{eq:87} and have proved Proposition~\bref{lem4}. \qed

\section{Hamiltonian formalism}
\label{sec:hamilt-form}

In this appendix, we will analyze how the results discussed so far
appear in the particular case of an Hamiltonian action where the
surface $S$ is a closed surface inside the space-like hyperplane
$\Sigma_t$ defined by constant $t$.

We follow closely the conventions and use results of
\cite{Henneaux:1992ig} for the Hamiltonian formalism. The Hamiltonian
action is first order in time derivatives and given by
\begin{equation}
S_H[z,\lambda] =\int \cL_H= \int dt d^{n-1}x \, (\dot z^A a_A - h -
\lambda^a \gamma_a )\,, \label{eqApp:1}
\end{equation}
where we assume that we have Darboux coordinates: $z^A=(\phi^\alpha,
\pi_\alpha)$ and $\dot z^A a_A=\dot \phi^\alpha \pi_\alpha$. It
follows that $\sigma_{AB}=\partial_A a_B-\partial_B a_A$ is the
constant symplectic matrix with $\sigma^{AB} \sigma_{BC} = \delta^A_C$
and $d^{n-1}x\equiv (d^{n-1}x)_0$. We assume for simplicity that the
constraints $\gamma_a$ are first class, irreducible and time
independent. In the following we shall use a local ``Poisson'' bracket
with spatial Euler-Lagrange derivatives for spatial $n-1$ forms $\hat
g= g\, d^{n-1}x$,
\begin{equation}
\{ \hat g_1,\hat g_2\} = 
\varQ{g_1}{z^A}\sigma^{AB}\varQ{g_2}{z^B}\,  d^{n-1}x.
\end{equation}
If $\tilde d_H$ denotes the spatial exterior derivative, this 
bracket defines a Lie bracket in the space $H^{n-1}(\tilde d_H)$,
i.e., in the space of equivalence classes of local functions modulo
spatial divergences, see e.g.~\cite{Barnich:1996mr}. 

Similarly, the Hamiltonian vector fields associated to an $n-1$ form $\hat
h = h \, d^{n-1}x$
\begin{eqnarray}
\stackrel{\leftarrow}{\delta_{\hat h}}\,(\cdot) =  \QS{}{z^A_{(i)}} (\cdot)\,
\sigma^{AB}\d_{(i)}\varQ{h}{z^B}=\{\cdot,\hat h\}_{alt},\\
\stackrel{\rightarrow}{\delta_{\hat h}}\,(\cdot) =
\d_{(i)}\varQ{h}{z^B}\sigma^{BA} \QS{}{z^A_{(i)}} (\cdot)\,
=\{\hat h,\cdot\}_{alt},\\
\end{eqnarray}
only depend on the class $[\hat h] \in H^{n-1}(\tilde d_H)$. Here
$(i)$ is a multi-index denoting the spatial derivatives, over which we
freely sum. The combinatorial factor needed to take the symmetry
properties of the derivatives into account is included in
$\frac{\partial^S}{\partial z^A_{(i)}}$. If we denote by $\hat
\gamma_a = \gamma_a \, d^{n-1}x$ and $\hat h_E =\hat h +\lambda^a
\hat\gamma_a$, an irreducible generating set of gauge transformations
for \eqref{eqApp:1} is given by
\begin{eqnarray}
  {\delta_{f}} z^A &=
  &\{z^A,\hat \gamma_a f^a\}_{alt},\label{cang1}\\
  {\delta_{f}}\lambda^a &=& \frac{D f^a}{Dt} +
\{f^a,\hat h_E\}_{alt} +\cC_{bc}^a(f^b,\lambda^c)-\cV_b^a(f^b),\label{cang2}
\end{eqnarray}
where the arbitrary gauge parameters $f^a$ may depend on $x^\mu$, the
Lagrange multipliers and their derivatives as well as the canonical
variables and their spatial derivatives and
\begin{eqnarray}
 &&\frac{D}{Dt} = \Q{}{t} + \dot\lambda^a\Q{}{\lambda^a} + \ddot \lambda^a
 \Q{}{\dot\lambda^a} + \dots,\\
  && \{ \gamma_a, \hat \gamma_b \lambda^b\}_{alt} =
  \cC^{+c}_{ab}(\gamma_c,\lambda^b), \\
  &&\{\gamma_a,\hat h\}_{alt} = -\cV_a^{+b}(\gamma_b).
\end{eqnarray}
For later use, we define $[f_1,f_2]^a_H=\cC_{bc}^a(f^b_1,f_2^c)$ and
assume the $f$'s to be independent jet-coordinates. 

Let us denote the set of fields collectively by $\phi^i = \{z^A,
\lambda^a\}$. In order to construct the surface charges, we first have
to compute the current $S_f$ defined according to \eqref{1.3},
\eqref{sec1cur}:
\begin{eqnarray}
R^i_\alpha (f^\alpha) \varQ{\cL_H}{\phi^i} &=&
\Big[\sigma^{AB}\frac{\delta(\gamma_a f^a )}{\delta z^B}(\sigma_{AC} \dot z^C -
\varQ{h}{z^A} - \varQ{\lambda^b \gamma_b}{z^A})\nonumber\\
&& \quad + (\frac{D f^a}{Dt} + \{f^a,\hat h_E\}_{alt}+
\cC_{bc}^a(f^b,\lambda^c)-V_b^a(f^b))\, (-\gamma_a)\Big]d^nx\nonumber\\
&\hspace*{-3cm}=& \hspace*{-1.5cm}
\Big[-
\frac{d}{dt}(\gamma_a f^a ) +
\d_k\big(V^k_B[\dot z^B - \sigma^{BA}\varQ{h_E}{z^A},\gamma_a f^a
]
+j^{ka}_b(\gamma_a, f^b)\big)\Big]d^nx.\label{eqApp:5}
\end{eqnarray}
Here, the current $j^{ka}_b(\gamma_a,f^b)$ is determined in terms of the
Hamiltonian structure operators through the formula
\begin{eqnarray}
\d_k j^{ka}_b(\gamma_a,f^b) &=& \gamma_a \cV^a_b(f^b) - f^b
\cV^{+a}_b(\gamma_a) \nonumber\\
&&\quad - \gamma_c \cC^c_{ab}(f^a,\lambda^b) +
f^a\cC^{+c}_{ab}(\gamma_c,\lambda^b),
\end{eqnarray}
while $V^i_A(Q^A,g)=\d_{(j)} (Q^A\varQ{g}{z^A_{(j)i}} )$. We have
\begin{eqnarray}
\partial_{(k)}Q^A\frac{\partial g}{\partial z^A_{(k)}} =
Q^A\vddl{g }{z^A} +\d_i V^i_A(Q^A,g),\label{eq:99}
\end{eqnarray}
if the function $g$ does not involve time
derivatives of $z^A$. In other words, $V^i_A(Q^A,g)$
coincides with the components of the $n-2$-form $I^{n-1}_Q(g d^{n-1}x)$ as
defined in \eqref{phihomotopy}, \eqref{cc2} with $\phi^i$ replaced by
$z^A$ and $n$ replaced by $n-1$, i.e., for spatial forms
with no time derivatives on $z^A$.

The weakly vanishing Noether current $S^\mu_f$ is thus given by
\begin{eqnarray}
S^0_f &=& -  \gamma_a f^a,\label{sham0}\\
S^k_f &=& V^k_B[\dot z^B - \sigma^{BA}\varQ{h_E}{z^A},\gamma_a f^a
]+j^{ka}_b(\gamma_a, f^b).\label{shami}
\end{eqnarray}

Note that $k^{[0i]}_f[\dv\phi,\phi]$, which is the relevant part of
the $n-2$ form $k_f$ at constant time, only involves the canonical
variables $\dv z^A,z^A$ and the gauge parameters $f^a$, but
not the Lagrange multipliers $\lambda^a$ nor their variations, $\dv
\lambda^a$. This is so because $S^0_f$ does not involve
$\lambda^a$ while the terms in
$S^k_f$ with time derivatives involve only time derivatives of $z^a$
and no Lagrange multipliers:
\begin{eqnarray}
k^{[0i]}_f[\dv\phi,\phi]=k^{[0i]}_f[\dv z,z]\label{eq:53}.
\end{eqnarray}
More precisely, it follows from \eqref{eq:57}, \eqref{sham0},
\eqref{shami} and  that
\begin{eqnarray}
  \label{eq:14}
  k^{[0i]}_f[\dv z,z]=\frac{|k|+1}{|k|+2}
\partial_{(k)}[\dv z^A \varQ{(-  \gamma_a
f^a)}{z^A_{(k)i}}-\dv z^A\varQ{V^i_B[\dot z^B,\gamma_a f^a
]}{z^A_{(k)0}}].
\end{eqnarray}
Equation \eqref{eA8} then allows one to show that $\varQ{V^i_B[\dot
z^B,\gamma_a f^a
]}{z^A_{(k)0}}=\frac{1}{|k|+1}\varQ{(\gamma_a f^a)
}{z^A_{(k)i}}$ so that the terms nicely combine to give
\begin{eqnarray}
  \label{eq:37}
  k^{[0i]}_f[\dv z,z]=-V^i_A[\dv z^A,\gamma_a f^a].
\end{eqnarray}
Let $d\sigma_i=2(d^{n-2}x)_{0i}$. For $S$ a closed surface inside the
hyperplane $\Sigma_t$ defined by constant $t$, the surface charge one
form is given by
\begin{eqnarray}
  \label{eq:101}
  \ndelta Q_f[\dv z]=\int_S  k^{[0i]}_f[\dv z]d\sigma_i.
\end{eqnarray}
Taking into account \eqref{eq:99}, we thus recover the following
result from the Hamiltonian approach \cite{Regge:1974zd}:
\begin{prop}
  In the context of the Hamiltonian formalism, the surface charge
  1-forms at constant time do not depend on the Lagrange multipliers
  and are given by the boundary terms that arise when converting the
  variation of minus the constraints smeared with gauge parameters
  into an Euler-Lagrange derivative contracted with the
  undifferentiated variation of the canonical variables,
  \begin{eqnarray}
    \label{eq:102}
    -\dv^z(\gamma_af^a)=-\dv z^A\vddl{\gamma_a
      f^a}{z^A}+\d_ik^{[0i]}_f[\dv z,z]. 
  \end{eqnarray}
\end{prop}

In addition, because of the simple way time derivatives enter into the
Hamiltonian action $\cL_H$, we have for all $Q^i_1,Q^i_2$,
\begin{eqnarray}
  \label{eq:42}
  W^0_{\vddl{\cL_H}{\phi}}[Q_1,Q_2]&=&-\sigma_{AB}Q_1^AQ_2^B\,,\\
T^{0}_{R_f}[\dv\phi, \vddl{\cL_H}{\phi}]&=&0\,,\
 E^{0i}_{\cL_H}[\dv\phi,\dv\phi]=0\,.\label{eq:42c}
\end{eqnarray}
Note that the last relation follows from our assumption that we are
using Darboux coordinates. 
As a consequence of the first relation, we then also have
\begin{eqnarray}
  \label{eq:106}
  W^0_{\vddl{\cL_H}{\phi}}[\dv\phi,R_f]\, d^{n-1}x
&=&-\dv z^A\vddl{(\hat \gamma_a
f^a)}{z^A}\,,\\ W^0_{\vddl{\cL_H}{\phi}}[R_{f_1},R_{f_2}]\,
d^{n-1}x&=&
\{\hat\gamma_a
f^a_1,\hat \gamma_b f^b_2\}\,.
\end{eqnarray}

Let us now analyze how we can recover the results of
\cite{Brown:1986ed} from the present perspective.  Suppose that
$\Sigma_t$ has as unique boundary $S^{\infty,t}$, assume that the
fields and gauge parameters belong to the spaces $\cF$ and $\cA_\cF$
defined ``off-shell'' (cf.~remark 2 of
subsection~\bref{sec:asympt-symm-algebra}) and that
\begin{eqnarray}
  \label{eq:44}
  T^0_{R_g}[\dv\phi,\vddl{\cL}{\phi}]=0. 
\end{eqnarray}
We can define the functionals
\begin{eqnarray}
  \label{eq:43}
 G[\Phi,\bar\Phi]=-\int_\Sigma S_g+ \cQ[\Phi,\bar\Phi],
\end{eqnarray}
i.e.,  the Noether charges associated with the weakly vanishing Noether
current $n-1$-forms $S_g$, ``improved'' by the boundary term 
$\cQ[\Phi,\bar\Phi]$. In the
Hamiltonian formalism, $G[\Phi,\bar\Phi]=\int \hat\gamma_a g^a +
\cQ[\Phi,\bar\Phi]$.
It then follows from Stokes' theorem and \eqref{eq:29a} that
\begin{eqnarray}
  \label{eq:64}
  \dv^\phi G&=&-\int_{\Sigma_t}
  W_{\delta\cL/\delta\phi}[\dv\phi,R_g].
\end{eqnarray}
By analogy with the Hamiltonian analysis \cite{Regge:1974zd}, in which
we get \[\dv^\phi G=\int_{\Sigma_t} \dv z^A\vddl{(\hat \gamma_a
  g^a)}{z^A},\] we say that the functional $G$ is
differentiable. Because $R_{g}$ is by assumption tangent to $\cF$, we
say furthermore that $G$ is a differentiable generator.

We can define the covariant Poisson bracket of the functionals $G_i\equiv G
[\Phi_i,\bar\Phi_i]$, by 
\begin{eqnarray}
\{G_{1} ,G_{2} \}_c=-\delta_{R_{g_1}} G_2,\label{eq:8}
\end{eqnarray}
By applying $i_{R_{g_1}}$ to \eqref{eq:64}
in terms of $g_2$,
\begin{eqnarray}
  \label{eq:61a}
\{G_{1} ,G_{2} \}_c =\int_{\Sigma_t} W_{\delta\cL/\delta\phi}[R_{g_1},R_{g_2}].
\end{eqnarray}
In the Hamiltonian formalism, it coincides with the usual one,
\[\{G_{1},G_{2}\}_c=\{\int_{\Sigma_t} \hat\gamma_a
g^a_1,G_{2}\}_{alt}= \int_{\Sigma_t}\{\hat\gamma_a g^a_1,\hat\gamma_b
g^b_2\}.\] 
 
If we now assume that the integrability conditions for the gauge
parameters $g$ are satisfied without the need to vary the parameters
$g$, as usually done in the Hamiltonian formalism,
\begin{equation}
 \oint_{S^{\infty,t}} k_{\dv
g}[\dv \phi] = 0,\label{int_dvg}
\end{equation}
it follows that $\{G_{1} ,G_{2} \}_c$ is differentiable,
\begin{eqnarray}
  \label{eq:52}
  \dv^\phi \{G_{1} ,G_{2} \}_c
=\int_{\Sigma_t} W_{\delta\cL/\delta\phi}[\dv\phi,R_{[g_1,g_2]}].
\end{eqnarray}
 
\vspace*{.25cm}
 
\begin{minipage}{.90\textwidth}\footnotesize 
 
Indeed, applying $\dv^\phi$ to \eqref{eq:29a} and using~\eqref{eq:44}
we get 
\begin{equation*}
\int_{\Sigma_t} \dv^\phi W_{\delta\cL/\delta\phi}[\dv\phi,R_g] =
-\int_{S^{\infty,t}} \dv^\phi k_g[\dv \phi].
\end{equation*}
Using integrability~\eqref{eq:92a}, \eqref{int_dvg} and the off-shell
assumptions, the expression on the RHS vanishes so that 
\begin{equation}
\int_{\Sigma_t} \dv^\phi W_{\delta\cL/\delta\phi}[\dv\phi,R_g] =0.\label{eq:last}
\end{equation}
Using~\eqref{eq:comm2} and \eqref{eq:79}, we have
\begin{multline}
\dv^\phi \{G_1,G_2\}_c = \int_{\Sigma_t}\dv^\phi
W_{\delta\cL/\delta\phi}[R_{g_1},R_{g_2}]
= \int_{\Sigma_t}\Big(
W_{\delta\cL/\delta\phi}[\dv\phi,R_{[g_1,g_2]}]+i_{R_{g_1}}
\dv^\phi i_{R_{g_2}}W_{\delta\cL/\delta\phi}\\ - i_{R_{g_2}}
\dv^\phi i_{R_{g_1}}W_{\delta\cL/\delta\phi}-\dH
i_{R_{g_1}}i_{R_{g_2}}\dv^\phi E_\cL\Big), \nonumber
\end{multline}
where the second and third terms on the RHS vanish as a consequence of
\eqref{eq:last}, while the last one vanishes on account of the
off-shell version of \eqref{eq:109}.
\end{minipage}
 
\vspace*{.25cm}
 
\noindent This is the theorem proved in \cite{Brown:1986ed} in the
Hamiltonian framework, where \[\dv^\phi \{G_{1},G_{2}\}_c=\int \dv
z^A\vddl{\hat \gamma_a[g_1,g_2]^a_H}{z^A}.\]

\section{General relativity}
\label{app:E}
We start from the Einstein-Hilbert action with cosmological constant
$\Lambda$
\begin{eqnarray}
  \label{eq:56}
  S[g]=\int\cL^{EH}= \int d^nx\, \frac{\sqrt{|g|}}{16\pi
  G}(R-2\Lambda).
\end{eqnarray}
A generating set of gauge transformations is given by
\begin{eqnarray}
  \label{eq:6}
  \delta_\xi g_{\mu\nu}=\xi^\rho\partial_\rho
  g_{\mu\nu}+\partial_\mu\xi^\rho g_{\rho\nu}+\partial_\nu\xi^\rho
  g_{\mu\rho}.
\end{eqnarray}
Reducibility parameters at $g$ are thus given by Killing vectors
of $g$. The weakly vanishing Noether current~\eqref{sec1cur} is
given by
\begin{eqnarray}
  \label{eq:15a}
  S^\mu_\xi[\varQ{L^{EH}}{g}]=2\varQ{L^{EH}}{g_{\mu\nu}}\xi_\nu=
\frac{\sqrt{|g|}}{8\pi G}(-G^{\mu\nu}-\Lambda g^{\mu\nu})\xi_\nu.
\end{eqnarray}
An explicit expressions for $k_\xi = I^{n-1}_{\dv g}S_\xi$
using~\eqref{phihomotopy} has been originally derived in
\cite{Barnich:2001jy}\footnote{Note however that we have changed
  conventions, which are here taken to be those of MTW
  \cite{Misner:1970aa} and introduced an overall minus sign in the
  definition.} and compared to other proposals in the literature. We
point out here that $k_\xi = I^{n-1}_{\dv g}S_\xi$ can also be written
in the compact form
\begin{equation}
k_\xi[\dv g] = \frac{2}{3}(d^{n-2}x)_{\mu\nu}P^{\mu\delta\nu
\gamma \alpha\beta}(2 D_\gamma \dv g_{\alpha\beta}\xi_\delta - \dv
g_{\alpha\beta} D_\gamma \xi_\delta),\label{ch_grav1}
\end{equation} 
where
\begin{eqnarray}
 P^{\mu\nu\alpha\beta\gamma
\delta}&=&\frac{\d^S}{\d g_{\gamma \delta,\alpha\beta}} \left(
\varQ{L^{EH}}{g_{\mu\nu}} \right)\\
&=&
\frac{\sqrt{-g}}{32\pi G}
\big( g^{\mu\nu} g^{\gamma(\alpha} g^{\beta)\delta} +
g^{\mu(\gamma}g^{\delta)\nu} g^{\alpha\beta} + g^{\mu(\alpha}
g^{\beta)\nu}g^{\gamma\delta}
\nonumber \\
&& - g^{\mu\nu} g^{\gamma\delta} g^{\alpha\beta}  -
g^{\mu(\gamma}g^{\delta)(\alpha}g^{\beta)\nu}  -g^{\mu(\alpha}
g^{\beta)(\gamma}g^{\delta)\nu} \big).
\end{eqnarray}
The tensor density $P^{\mu\nu\alpha\beta\gamma \delta}$ is related to
the supermetric defined in~\cite{Dewitt:1967yk}, which has the
symmetries of the Riemann tensor, through
\begin{eqnarray}
  \label{eq:51}
\frac{2}{n-2}g_{\mu\nu}P^{\mu\nu\alpha\beta\gamma \delta}= G^{\alpha
  \beta\gamma\delta},
\end{eqnarray}
where
\begin{eqnarray}
G^{\alpha \beta\gamma\delta} &=&  \frac{\d^S L^{EH}}{\d g_{\gamma
\delta,\alpha\beta}}\\ &=&\frac{\sqrt{-g}}{16\pi G} \big(
\half g^{\alpha\gamma} g^{\beta\delta} + \half g^{\alpha\delta}
g^{\beta\gamma} -  g^{\alpha\beta} g^{\gamma\delta} \big)\label{deWitt}.
\end{eqnarray}
The tensor density $P^{\mu\nu\alpha\beta\gamma \delta}$ itself is
symmetric in the pair of indices $\mu\nu$, $\alpha\beta$ and $\gamma
\delta$ and the total symmetrization of any three indices is zero. The
symmetries of these tensors are thus summarized by the Young tableaux
\begin{equation}
G^{\alpha \beta\gamma\delta} \sim
\begin{picture}(28,25)(0,0)
\multiframe(0,5.75)(11.5,0){2}(11,11){$\alpha$}{$\beta$}
\multiframe(0,-5.75)(11.5,0){2}(11,11){$\gamma$}{$\delta$}
\end{picture},
\qquad P^{\mu \nu \alpha\beta\gamma\delta} \sim
\begin{picture}(28,25)(0,0)
\multiframe(0,11.5)(11.5,0){2}(11,11){$\mu$}{$\nu$}
\multiframe(0,0)(11.5,0){2}(11,11){$\alpha$}{$\beta$}
\multiframe(0,-11.5)(11.5,0){2}(11,11){$\gamma$}{$\delta$}
\end{picture}.\vspace{11pt}\label{pmunu}
\end{equation}
Note that in the context of covariant phase space methods
\cite{Wald:1999wa}, a similar tensor density
$P^{\mu\nu\alpha\beta\gamma \delta}$ is defined, which lacks however
the above symmetry properties. In order to compare further with the
formulae obtained in that approach, it is convenient to rewrite
$k_\xi[\dv g]$ as
\begin{equation}
k_\xi[\dv \phi] =  -\dv k^K_{\cL^{EH},\xi}
+k^K_{\cL^{EH},\dv\xi} +i_\xi I^{n}_{\dv
\phi}\cL^{EH}-E_{\cL^{EH}}[\cL_\xi \phi,\dv \phi],\label{k_diff2}
\end{equation}
where 
\begin{eqnarray}
k^K_{\cL^{EH},\xi} = \frac{\sqrt{-g}}{16\pi G} (D^\mu \xi^\nu -
D^\nu
\xi^\mu )(d^{n-2}x)_{\mu\nu},\label{Komar}
\end{eqnarray}
is the Komar integral, 
\begin{eqnarray}
  \label{eq:41}
  I_{\dv g}^{n}\cL^{EH}[\dv g] = \frac{\sqrt{-g}}{16\pi G}
(g^{\mu\alpha} D^\beta \dv g_{\alpha\beta} - g^{\alpha\beta} D^\mu
\dv g_{\alpha\beta} ) (d^{n-1}x)_\mu,\label{Theta_t}
\end{eqnarray}
and the additional term
\begin{equation}
E_{\cL^{EH}}[\cL_\xi g,\dv g] = \frac{\sqrt{-g}}{16\pi G}
(\half g^{\mu\alpha}\dv g_{\alpha\beta} (D^\beta \xi^\nu + D^\nu
\xi^\beta) - (\mu \leftrightarrow \nu) )
(d^{n-2}x)_{\mu\nu},\label{suppl}
\end{equation}
vanishes for exact Killing vectors of $g$, but not necessarily for
asymptotic ones. 

Explicitly, the quantities involved in $\dH
k_\xi[\dv g]$ in~\eqref{eq:29a}
are given by 
\begin{eqnarray}\begin{aligned}
W_{\varQ{\cL^{EH}}{\phi}}[\dv g,\cL_\xi g] &=
P^{\mu\delta\beta\gamma\varepsilon\zeta}\Big( \dv g
_{\beta\gamma}\nabla_\delta \cL_\xi g_{\varepsilon\zeta}- \cL_\xi
g_{\beta\gamma}\nabla_\delta \dv g_{\varepsilon\zeta}
 \Big) (d^{n-1}x)_\mu,\\
T_{\cL_\xi g}[\dv g , \varQ{\cL^{EH}}{g}]&=
 \dv g_{\alpha\beta}\varQ{L^{EH}}{g_{\alpha\beta}}\xi^\mu
(d^{n-1}x)_\mu.\end{aligned}\label{dHgrav}
\end{eqnarray}
It follows that all conditions are satisfied to proceed with an
off-shell analysis in the asymptotic context, cf.~remark 2 of
subsection~\bref{sec:asympt-symm-algebra}. 

\addcontentsline{toc}{section}{References}

\providecommand{\href}[2]{#2}\begingroup\raggedright\endgroup

\end{document}